\documentclass[aps,prd,preprintnumbers,superscriptaddress,showpacs,nofootinbib,11pt,floatfix]{revtex4}
\usepackage{graphicx}
\usepackage{dcolumn}
\usepackage{bm}
\usepackage{amsmath}
\usepackage{amssymb}

\def\beq{\begin{equation}}
\def\eeq{\end{equation}}
\def\[{\left[}
\def\]{\right]}

\def\gsim{\lower.7ex\hbox{$\;\stackrel{\textstyle>}{\sim}\;$}}
\def\lsim{\lower.7ex\hbox{$\;\stackrel{\textstyle<}{\sim}\;$}}

\begin{document}
\begin{flushright}
{SHEP-08-27}\\
{DCP-08-03}\\
\end{flushright}
\title{Yukawa textures and charged Higgs boson phenomenology in the type-III two-Higgs-doublet model}
\author{J. L. D\'{\i}az-Cruz}
\email{jldiaz@fcfm.buap.mx} \affiliation{Fac. de Cs.
F\'{\i}sico-Matem\'aticas, BUAP. Apdo. Postal 1364, C.P. 72000
Puebla, Pue., M\'exico, and Dual C-P Institute of High Energy Physics, M\'exico.}
\author{J. Hern\' andez--S\' anchez}
\email{jaimeh@ece.buap.mx} \affiliation{Fac. de Cs. de la
Electr\'onica, BUAP, Av. San Claudio y 18 Sur, C.P. 72500 Puebla, Pue.,
M\'exico, and Dual C-P Institute of High Energy Physics, M\'exico.}
\author{S. Moretti}
\email{stefano@soton.ac.uk} \affiliation{ School of  Physics and Astronomy,
University of Southampton, Highfield, Southampton SO17 1BJ, UK.}
\author{R. Noriega-Papaqui}
\email{rnoriega@uaeh.edu.mx}
\affiliation{Centro de Investigaci\'on en Matem\'aticas,
Universidad Aut\'onoma del Estado de Hidalgo,
Carr. Pachuca-Tulancingo Km. 4.5, C.P. 42184, Pachuca, Hgo., M\'exico, and Dual C-P Institute of High Energy Physics, M\'exico.}
\author{A. Rosado}
\email{rosado@sirio.ifuap.buap.mx} \affiliation{ Instituto de
F\'{\i}sica, BUAP. Apdo. Postal J-48, C.P. 72570 Puebla, Pue.,
M\'exico.}
\date{\today}
\begin{abstract}
\noindent
We discuss the implications of assuming a four-zero Yukawa texture
for the properties of the charged Higgs boson within the context of
the general 2-Higgs Doublet Model of Type III. We begin by presenting a
detailed analysis of the charged Higgs boson
couplings with heavy quarks and the resulting pattern for its
decays. The production of charged Higgs bosons is also sensitive to
the modifications of its couplings, so that we also evaluate the resulting
effects on the top decay $t \to b H^{+}$ as well as on `direct'
$c\bar{b}\to H^++c.~c.$ and `indirect' $q\bar q,gg\to \bar t b H^++c.~c.$
production. Significant scope exists at the
Large Hadron Collider for several $H^\pm$ production and decay channels
combined to enable one to distinguish between such a model and alternative
2-Higgs doublet scenarios.
\end{abstract}
\pacs{12.60.Cn,12.60.Fr,11.30.Er}
\maketitle

\newpage

\section{Introduction}

Detecting a charged Higgs boson during the imminent Large Hadron
Collider (LHC) experimental running would constitute a clear
evidence of physics beyond the Standard Model (SM) \cite{stanmod}.
Charged Higgs bosons appear in many well motivated extensions of the
SM, whose phenomenology has been widely studied over the years
\cite{kanehunt,susyhix1,susyhix2}. In particular, 2-Higgs Doublet
Models (2HDMs), in both Supersymmetry (SUSY) and non-SUSY versions
\cite{LorenzoDiazCruz:2008zz,Djouadi:2005gj}, can be considered as a
prototype of a Higgs sector that includes a charged Higgs boson
($H^\pm$). It is expected that the LHC will allow us to test the
mechanism of Electro-Weak Symmetry Breaking (EWSB) and, in
particular, to probe the properties of charged Higgs bosons, which
represent a unique probe of a weakly-interacting theory, as is the
case of the Minimal Supersymmetric Standard Model (MSSM)
\cite{LorenzoDiazCruz:2008zz} and general 2HDMs of Type I, II, III
and IV (2HDM-I, 2HDM-I, 2HDM-III and 2HDM-IV) \cite{Barger:1989fj},
 or whether strongly-interacting
scenarios are instead realized, like in the old Technicolor models or
similar ones discussed
more recently \cite{stronghix}. Ultimately, while many analyses in
this direction can be carried out at the LHC, it will be a future
International Linear Collider (ILC) \cite{ILC} or Compact LInear Collider (CLIC)
\cite{CLIC}
which will have the definite word about exactly which mechanism of mass generation
and which realization of it occurs in Nature.

The 2HDM-II has been quite attractive to date, in part because it coincides
with the Higgs sector of the MSSM, wherein each Higgs doublet couples
to the $u$- or $d$-type fermions separately\footnote{Notice that there exist
significant
differences between the 2HDM-II and MSSM though, when it comes to their
mass/coupling configurations and possible Higgs signals \cite{Kanemura:2009mk}.}.
However, this is only valid at
tree-level \cite{Babu-Kolda}. When radiative effects are included, it
turns out that the MSSM Higgs sector corresponds to the most general
version of the 2HDM, namely the 2HDM-III, whereby both Higgs fields
couple to both quarks and leptons. Thus, we can consider the
2HDM-III as a generic description of physics at a higher scale (of
order TeV or maybe even higher), whose low energy imprints are
reflected in the Yukawa coupling structure. With this idea in
mind, some of us have presented a detailed study of the 2HDM-III Yukawa
Lagrangian \cite{ourthdm3a}, under the assumption of a specific
texture pattern \cite{Fritzsch:2002ga}, which generalizes the
original model of Ref.~\cite{cheng-sher}. Phenomenological
implications of this model for the neutral Higgs sector, including
Lepton Flavour Violation (LFV) and/or
Flavour Changing Neutral Currents (FCNCs) have been presented in a
previous work
\cite{ourthdm3b}. Here, we are interested in extending such an approach
to investigate charged Higgs boson phenomenology: namely, we want to
study the implications of this Yukawa texture for the charged Higgs
boson properties (masses and couplings) and discuss in detail the resulting pattern of
charged Higgs boson decays and main production reactions at the LHC.

Decays of charged Higgs bosons have been studied in the literature,
including the radiative modes $W^{\pm}\gamma, W^{\pm}Z^0$
\cite{hcdecay}, mostly within the context of the 2HDM-II or its SUSY
incarnation (i.e., the MSSM), but also by
using an effective Lagrangian extension of the
2HDM \cite{ourpaper} and, more recently, within an extension of the
MSSM with one Complex Higgs Triplet (MSSM+1CHT) \cite{ourtriplets,
Barradas-Guevara:2004qi}. Charged Higgs boson production at hadron
colliders was studied long ago \cite{ldcysampay} and, more recently,
systematic calculations of production processes at the LHC
have been presented \cite{newhcprod}.

Current bounds on the mass of a charged Higgs boson have been obtained at
Tevatron, by studying the top decay $t \to b \, H^+$, which already
eliminates large regions of the parameter space
\cite{Abulencia:2005jd}, whereas LEP2 bounds imply that, approximately,
$m_{H^{+}} > 80$ GeV \cite{lepbounds,partdat}, rather
model independently.
Concerning theoretical limits,
tree-level unitarity bounds on the 2HDM Higgs masses have been studied
in generic 2HDMs and in particular an upper limit for the charged
Higgs mass of
800 GeV or so can be obtained, according to  the results of
Ref.~\cite{unitarity}.

This paper  is organized as follows. In section
II, we discuss the Higgs-Yukawa sector of the 2HDM-III, in particular,
we derive the expressions for the charged Higgs boson couplings to
heavy fermions. Then, in section
III, we derive the expressions for the decays $H^+ \to f_i
\bar{f}_j$ and numerical results are presented for some 2HDM-III scenarios,
defined for phenomenological purposes. A discussion of the main
production mechanisms at the LHC is presented in  section IV. These
include the top decay $t \to b \, H^+$ as well as $s$-channel
production of charged Higgs bosons through
$c\bar{b}(\bar{c}b)$-fusion \cite{He:1998ie} and the multi-body
more  $q \bar{q}, \, gg \to t \bar{b} H^- + $ c.c. (charge conjugated). These
mechanisms depend crucially on the parameters of the underlying model and
large deviations should be expected in the 2HDM-III
with respect to the 2HDM-II. Actual LHC
event rates are given in section V. Finally, we summarize our results
and present the conclusions in section VI. Notice that in
carrying out this plan, unlike other references \cite{chinos,colombianos},
where the 2HDM-II and the 2HDM-III appear as different structures, we
shall consider here that, under certain limits, the 2HDM-III reduces to
the 2HDM-II and, therefore, that the properties of the charged Higgs bosons
change continuously from one model to the other.

\section{The Charged Higgs boson Lagrangian and the fermionic couplings}

We shall follow Refs.~\cite{ourthdm3a, ourthdm3b}, where a
specific four-zero texture has been implemented for the Yukawa
matrices within the 2HDM-III. This allows one to express the couplings
of the neutral and charged Higgs bosons in terms of the fermion
masses, Cabibbo-Kobayashi-Maskawa
(CKM) mixing angles and certain dimensionless parameters,
which are to be bounded by current experimental constraints. Thus,
in order to derive the interactions of the charged Higgs boson, the
Yukawa Lagrangian is written as follows: \beq {\cal{L}}_{Y} =
Y^{u}_1\bar{Q}_L {\tilde \Phi_{1}} u_{R} +
                   Y^{u}_2 \bar{Q}_L {\tilde \Phi_{2}} u_{R} +
Y^{d}_1\bar{Q}_L \Phi_{1} d_{R} + Y^{d}_2 \bar{Q}_L\Phi_{2}d_{R},
\label{lagquarks} \eeq \noindent where $\Phi_{1,2}=(\phi^+_{1,2},
\phi^0_{1,2})^T$ refer to the two Higgs doublets, ${\tilde
\Phi_{1,2}}=i \sigma_{2}\Phi_{1,2}^* $, $Q_{L}$ denotes the
left-handed fermion doublet, $u_{R} $ and $d_{R}$ are the
right-handed fermions singlets and, finally, $Y_{1,2}^{u,d}$ denote the
$(3 \times 3)$ Yukawa matrices. Similarly, one can
write the corresponding Lagrangian for leptons.

After spontaneous EWSB  and including the
diagonalizing matrices for quarks and Higgs bosons\footnote{The
details of both diagonalizations are presented in
Ref.~\cite{ourthdm3a}}, the interactions of the charge Higgs boson
$H^+$ with quark pairs acquire the following form:
\begin{eqnarray}
\label{QQH} {\cal{L}}^{\bar{q}_i q_j H^+} & = & \frac{g}{2
\sqrt{2} M_W} \sum_{l=1}^{3} \bar{u}_i \left\{  (V_{\rm CKM})_{il}
\left[ \tan \beta \, m_{d_{l}} \, \delta_{lj} -\sec \beta
\left(\frac{\sqrt{2} M_W}{g}\right) \left(
\tilde{Y}^d_2\right)_{lj}  \right] \right.
\nonumber \\
& & + \left[ \cot \beta \, m_{u_{i}} \, \delta_{il} -\csc \beta
\left(\frac{\sqrt{2} M_W}{g}\right) \left(
\tilde{Y}^u_1\right)_{il}^{\dagger}  \right]
(V_{\rm CKM})_{lj} \nonumber \\
& & + (V_{\rm CKM})_{il} \left[ \tan \beta \, m_{d_{l}} \, \delta_{lj}
-\sec \beta  \left(\frac{\sqrt{2} M_W}{g}\right) \left(
\tilde{Y}^d_2\right)_{lj}  \right] \gamma^{5}\\
& & - \left. \left[ \cot \beta \, m_{u_{i}} \, \delta_{il} -\csc
\beta  \left(\frac{\sqrt{2} M_W}{g}\right) \left(
\tilde{Y}^u_1\right)_{il}^{\dagger}  \right] (V_{\rm CKM})_{lj} \,
\gamma^{5} \right\} \, d_{j} \, H^{+},   \nonumber
\end{eqnarray}
where $V_{\rm CKM}$ denotes the mixing matrices of the quark
sector (and similarly for the leptons). The term proportional to
$\delta_{ij}$ corresponds to the contribution that would arise
within the 2HDM-II, while the terms proportional to
$\tilde{Y}_2^d$ and $\tilde{Y}_1^u$ denote the new contributions
from the 2HDM-III. These contributions, depend on the rotated
matrices: $\tilde{Y}_n^{q} = O_q^{T}\, P_q\, Y^{q}_n \,
P_q^\dagger \, O_q$ ($n=1$ when $q=u$, and $n=2$ when $q=d$ ),
where $O_q$ is the diagonalizing matrix, while $P_q$ includes the
phases of the Yukawa matrix. In order to evaluate
$\tilde{Y}^{q}_n$ we shall consider that all Yukawa matrices have
the Hermitian four-zero texture form~\cite{Fritzsch:2002ga}, and
the quark masses have the same form, which are given by:
\begin{equation} M^q= \left( \begin{array}{ccc}
0 & C_{q} & 0 \\
C_{q}^* & \tilde{B}_{q} & B_{q} \\
0 & B_{q}^*  & A_{q}
\end{array}\right)  \qquad
(q = u, d) .
\end{equation}
This is called a four-zero texture because one assumes that the
Yukawa matrices are Hermitian, therefore each $u$ and $d$ type
Yukawa matrix contains two independent zeros. According to current
analyzes this type of texture satisfies the experimental
constraints and at the same time it permits to derive analytical
expressions for the Higgs boson fermion couplings.

\begin{figure}
\centering
\includegraphics[width=6in]{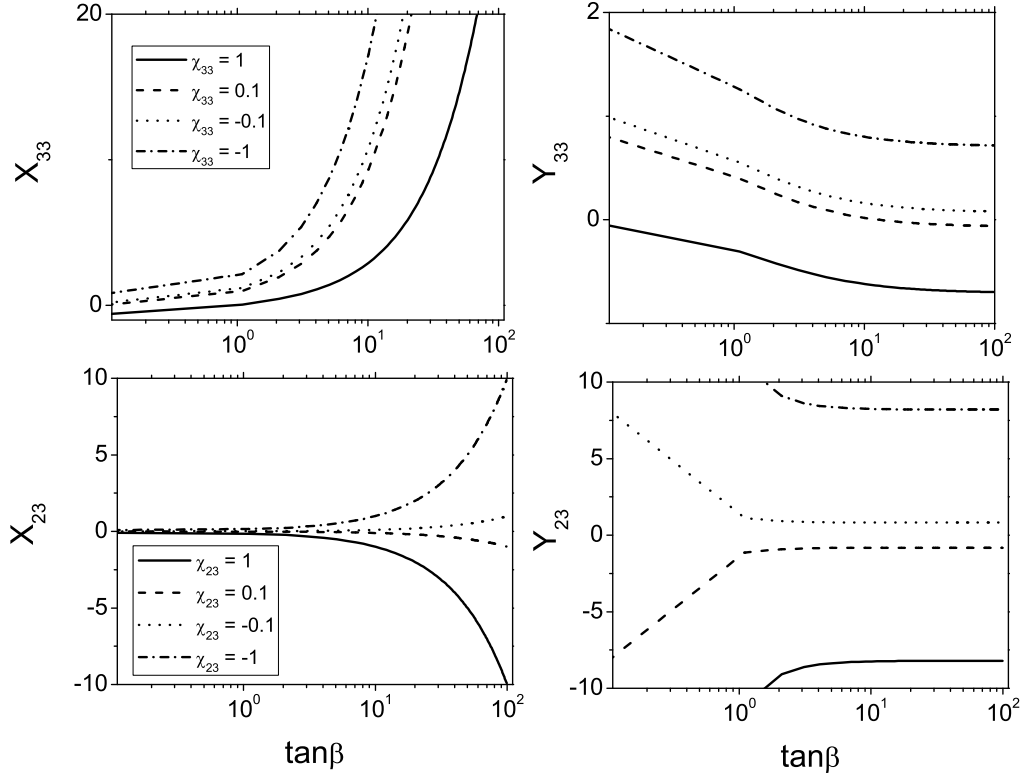}
\caption{The figure shows $X_{33}$, $Y_{33}$, $X_{23}$ and
$Y_{23}$ {\it vs.} $\tan \beta$, taking
$\tilde{\chi}_{3,3}^{u,d}=1$ (solid), $\tilde{\chi}_{3,3}^{u,d}=0.1$
(dashes), $\tilde{\chi}_{3,3}^{u,d}=-0.1$ (dots),
$\tilde{\chi}_{3,3}^{u,d}=-1$ (dashes-dots).} \label{fig:xy}
\end{figure}

To diagonalize $M^q$, we use the matrices $O_q$ and $P_q$, in the
following way \cite{Fritzsch:2002ga}:
\begin{equation}
\bar{M}^q = O^T_q \, P_q \, M^{q} \,P^{\dagger}_q \, O_q.
\label{masa-diagonal}
\end{equation}
Then, one can derive a better approximation for the product
$O_q^{T}\, P_q\, Y^{q}_n \, P_q^\dagger \, O_q$, expressing the
rotated matrix $\tilde {Y}^q_n$, in the form
\begin{equation}
\left[ \tilde{Y}_n^{q} \right]_{ij}
= \frac{\sqrt{m^q_i m^q_j}}{v} \, \left[\tilde{\chi}_{n}^q \right]_{ij}
=\frac{\sqrt{m^q_i m^q_j}}{v}\,\left[\chi_{n}^q \right]_{ij}  \, e^{i \vartheta^q_{ij}}.
\end{equation}
In order to perform our phenomenological study, we find it convenient
to rewrite the Lagrangian given in Eq.~(\ref{QQH})  in terms of the
coefficients $ \left[\tilde{\chi}_{n}^q \right]_{ij}$, as follows:
\begin{eqnarray}
\label{LCCH} {\cal{L}}^{q} & = & \frac{g}{2 \sqrt{2} M_W}
\sum_{l=1}^{3} \bar{u}_i \left\{  (V_{\rm CKM})_{il} \left[ \tan
\beta \, m_{d_{l}} \, \delta_{lj} -\frac{\sec \beta}{\sqrt{2} }
\,\sqrt{m_{d_l} m_{d_j} } \, \tilde{\chi}^d_{lj}  \right] \right.
\nonumber \\
& & + \left[ \cot \beta \, m_{u_{i}} \, \delta_{il}
  -\frac{\csc \beta}{\sqrt{2} }  \,\sqrt{m_{u_i} m_{u_l} } \, \tilde{\chi}^u_{il} \right]
  (V_{\rm CKM})_{lj} \nonumber \\
& & + (V_{\rm CKM})_{il} \left[ \tan \beta \, m_{d_{l}} \, \delta_{lj}
-\frac{\sec \beta}{\sqrt{2} }  \,\sqrt{m_{d_l} m_{d_j} } \, \tilde{\chi}^d_{lj}   \right] \gamma^{5}
 \\
& & - \left. \left[ \cot \beta \, m_{u_{i}} \, \delta_{il}
  -\frac{\csc \beta}{\sqrt{2} }  \,\sqrt{m_{u_i} m_{u_l} } \, \tilde{\chi}^u_{il}  \right]
  (V_{\rm CKM})_{lj} \, \gamma^{5} \right\} \, d_{j} \, H^{+},   \nonumber
\end{eqnarray}
where we have redefined $\left[ \tilde{\chi}_{1}^u \right]_{ij} =
\tilde{\chi}^u_{ij}$ and $\left[ \tilde{\chi}_{2}^d \right]_{ij} =
\tilde{\chi}^d_{ij}$. Then, from Eq.~(\ref{LCCH}), the couplings $\bar{u}_i d_j H^+$ and $u_i \bar{d}_j
H^-$ are given by:
\begin{eqnarray}
\label{coups1} g_{H^+\bar{u_i}d_j} &=& -\frac{ig}{ 2\sqrt{2} M_W}
(S_{i j} +P_{i j} \gamma_5), \quad g_{H^- u_i \bar{d_j}}= -\frac{ig
}{2\sqrt{2} M_W}  (S_{i j} -P_{i j} \gamma_5),
\end{eqnarray}
where $S_{i j}$ and $P_{i j}$ are defined as:
\begin{eqnarray}
\label{sp} S_{i j} & = & \sum_{l=1}^{3} (V_{\rm CKM})_{il} \bigg[
\tan \beta \, m_{d_{l}} \, \delta_{lj} -\frac{\sec \beta}{\sqrt{2}
}  \,\sqrt{m_{d_l} m_{d_j} } \, \tilde{\chi}^d_{lj}  \bigg]
\nonumber \\
& & + \bigg[ \cot \beta \, m_{u_{i}} \, \delta_{il}
  -\frac{\csc \beta}{\sqrt{2} }  \,\sqrt{m_{u_i} m_{u_l} } \, \tilde{\chi}^u_{il} \bigg]
  (V_{\rm CKM})_{lj}, \nonumber \\
P_{i j} & = & \sum_{l=1}^{3} (V_{\rm CKM})_{il} \bigg[ \tan \beta
\, m_{d_{l}} \, \delta_{lj}
-\frac{\sec \beta}{\sqrt{2} }  \,\sqrt{m_{d_l} m_{d_j} } \, \tilde{\chi}^d_{lj}   \bigg]  \\
& & - \bigg[ \cot \beta \, m_{u_{i}} \, \delta_{il}
  -\frac{\csc \beta}{\sqrt{2} }  \,\sqrt{m_{u_i} m_{u_l} } \, \tilde{\chi}^u_{il}  \bigg]
  (V_{\rm CKM})_{lj}. \nonumber
\end{eqnarray}
As it was discussed in Ref.~\cite{ourthdm3a}, most low-energy
processes imply weak bounds on the coefficients
$\tilde{\chi}^q_{ij}$, which turn out to be of $O(1)$. However, some
important constraints on $\tan \beta$ have started to appear, based
on $B$-physics \cite{Dudley:2009zi}. In order to discuss these
results we find convenient to generalize the notation of Ref.~\cite{Borzumati:1998nx} and define the couplings
$\bar{u}_i d_j H^+$ and $u_i \bar{d}_j H^-$ in terms of the
matrices $X_{ij}$, $Y_{ij}$ and $Z_{ij}$ (for leptons). In our case
these matrices are given by:
\begin{eqnarray}
X_{l j} & = &   \bigg[ \tan \beta \,
\delta_{lj} -\frac{\sec \beta}{\sqrt{2} }  \,\sqrt{\frac{m_{d_j}}{m_{d_l}}  }
\, \tilde{\chi}^d_{lj}  \bigg],
\nonumber \\
Y_{i l} & = &  \bigg[ \cot \beta \,  \delta_{il}
  -\frac{\csc \beta}{\sqrt{2} }  \,\sqrt{\frac{ m_{u_l}}{m_{u_i}} } \, \tilde{\chi}^u_{il}  \bigg] .
\end{eqnarray}
where $X_{lj}$ and $Y_{il}$ are related with $S_{ij}$ and $P_{ij}$
defined in the Eq.~(\ref{sp}) as follows:
\begin{eqnarray}
S_{i j} & = &  \sum_{l=1}^{3} \left[(V_{\rm CKM})_{il}  \,
m_{d_{l}} \, X_{lj} +  m_{u_{i}} \, Y_{il}  (V_{\rm CKM})_{lj} \right], \nonumber \\
P_{i j} & = &  \sum_{l=1}^{3} \left[(V_{\rm CKM})_{il}  \,
m_{d_{l}} \, X_{lj} -  m_{u_{i}} \, Y_{il}  (V_{\rm CKM})_{lj}
\right].
\end{eqnarray}
The 33 elements of these matrices reduce to the expressions for the
parameters X,Y,Z ($=X_{33},Y_{33},Z_{33}$) used in
Ref.\cite{Borzumati:1998nx}. Based on the analysis of $B \to X_s
\gamma$ \cite{Borzumati:1998nx, Xiao:2003ya}, it is claimed that $X
\leq 20$ and $Y \leq 1.7$ for $m_{H^+} > 250$ GeV, while for a
lighter charged Higgs boson mass, $m_{H^+} \sim 200$ GeV, one gets
$(X,Y) \leq (18,0.5)$. Fig.~\ref{fig:xy} shows the values of $(X,Y)$
as a function of $\tan \beta$ within our model. Thus, we find
important bounds: $|\chi_{33}^{u,d}| \lsim 1$ for $0.1 < \tan \beta
\leq 70$. Although in our model there are additional contributions
(for instance from $c$-quarks, which are proportional to $X_{23}$),
they are not relevant because the Wilson coefficients in the
analysis of $B \to X_s \gamma$ are functions of $m_{c}^2/M_W^2$ or
$m_{c}^2/m_{H^+}^2$ \cite{Deshpande:1987nr}, that is, negligible
 when compared to the leading $X_{33}$ effects, whose Wilson
coefficients depend on $m_t^2/M_W^2$ or $m_t^2/m_{H^+}^2$. Other
constraints on the charged Higgs mass and $\tan\beta$, based on
$\Delta a_{\mu}$, the $\rho$ parameter, as well as B-decays into the
tau lepton, can be obtained
\cite{BowserChao:1998yp,WahabElKaffas:2007xd}. For instance, as can
be read from Ref.\cite{Isidori:2007ed}, one has that the decay $B
\to \tau \nu$, implies a constraint such that for $m_{H^+}=200$
(300) GeV, values of $\tan\beta$ less than about 30 (50) are still
allowed, within MSSM or 2HDM-III: However, these constraints can
only be taken as estimates, as it is likely that they would be
modified for 2HDM-III. In summary, we find that low energy
constraints still allow to have $\tilde{\chi}^q_{ij}=
O(1)$\footnote{A more detailed analysis that includes the most
recent data is underway \cite{lorenzoetal}.}.
\section{\bf Decays of the charged Higgs boson}

Let us now discuss the decay modes of the charged Higgs boson
within our model. Hereafter, we shall refer to four benchmark scenarios,
namely. (i) {\bf Scenario A}:  $\tilde{\chi}^u_{ij}=1$,
$\tilde{\chi}^d_{ij}=1$; (ii) {\bf Scenario B}:
$\tilde{\chi}^u_{ij}=0.1$, $\tilde{\chi}^d_{ij}=1$; (iii) {\bf
Scenario C}: $\tilde{\chi}^u_{ij}=1$, $\tilde{\chi}^d_{ij}=0.1$;
(iv) {\bf Scenario D}: $\tilde{\chi}^u_{ij}=0.1$,
$\tilde{\chi}^d_{ij}=0.1$. We have performed the numerical analysis
of charged Higgs boson decays by taking $\tan\beta=0.1, \, 1, \, 15,
\, 70$ and varying the charged Higgs boson mass within the interval
100 GeV $\leq m_{H^{\pm}} \leq$ 1000 GeV, further fixing $m_{h^0}= 120$
GeV, $m_{A^0}=300$ GeV and the mixing angle at $\alpha = \pi /2$.

The condition $\frac{\Gamma_{H^+}}{m_{H^+}} < \frac{1}{2}$ in the
frame of the 2HDM-II implies $\frac{\Gamma_{H^+}}{m_{H^+}} \approx
\frac{3G_F m_t^2}{4\sqrt{2}\pi\tan\beta^2}$ which leads to $0.3
\lsim \tan\beta \lsim 130$. However, in the 2HDM-III we have that
$\frac{\Gamma_{H^+}}{m_{H^+}} \approx \frac{3G_F
m_t^2}{4\sqrt{2}\pi\tan\beta^2} \bigg(
\frac{1}{1-\frac{\tilde{\chi}^u_{33}}{\sqrt{2} \cos\beta}}\bigg)^2$,
we have checked numerically that this leads to $0.08 < \tan\beta <
200$ when $|\tilde{\chi}^u_{33}| \approx 1$ and  $0.3 < \tan\beta <
130$ as long as $|\tilde{\chi}^u_{33}| \to 0$ recovering the result
for the case of the 2HDM-II \cite{Chankowski:1999ta,Barger:1989fj}.
In this sense, if we consider the constraints imposed by the
perturbativity bound, a portion of the low $\tan\beta$ appearing in
some graphs would be excluded. However, we have decided to keep that
range both to show the behaviour of the quantities of interest, and
also because we have to keep in mind that such criteria
(perturbativity) should be taken as an order of magnitude
constraint.

The expressions for the charged Higgs boson decay widths $H^+ \to
u_i \bar{d}_j$ are of the form:
\begin{eqnarray}
\Gamma (H^+ \to u_i \bar{d}_j) &=& \frac{3 g^2}{32 \pi M_W^2 m_{H^+}^3 } \lambda^{1/2}(m_{H^+}^2, m_{u_i}^2, m_{d_j}^2) \nonumber\\
& & \times \, \bigg( \frac{1}{2}\bigg[ m_{H^+}^2-m_{u_i}^2-m_{d_j}^2\bigg] (S_{ij}^2+P_{ij}^2)- m_{u_i} m_{d_j}
(S_{ij}^2- P_{ij}^2 ) \bigg),
\end{eqnarray}
where $\lambda$ is the usual kinematic factor $\lambda(a,b,c)=
(a-b-c)^2-4bc$. When we replace $\tilde{\chi}_{ud} \to 0$, the
formulae of the decays width become those of the 2HDM-II: see, e.g.,
Ref.~\cite{kanehunt}. Furthermore, the expressions for the charged Higgs
boson decay widths of the bosonic modes  remain the same as in the
2HDM-II. Then the results for the Branching Ratios (BRs) are shown in
Figs.~2--8, and have the following characteristics.

\begin{figure}
\centering
\includegraphics[width=6in]{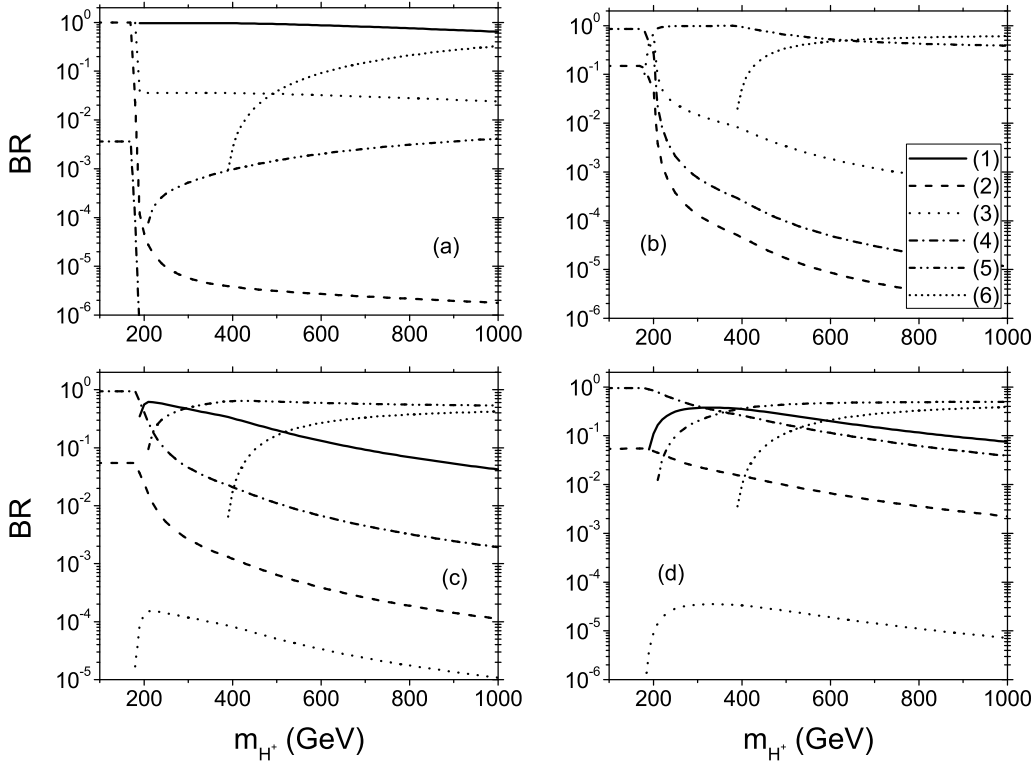}
\caption{The figure shows the BRs of the $H^+$ decaying into the
principal modes in Scenario A, taking $\tilde{\chi}_{ij}^{u}=1$,
$\tilde{\chi}_{ij}^{d}=1$, $m_{h^0} = 120$ GeV, $m_{A^0} = 300$
GeV and $\alpha = \pi/2$  for: (a) $\tan \beta=0.1$, (b) $\tan
\beta = 1$, (c) $\tan \beta = 15$, (d) $\tan \beta = 70$. The
lines  in each graph correspond to: (1) BR($H^+ \to t\bar{b} $),
(2) BR($H^+ \to c\bar{b}$), (3) BR($H^+ \to t\bar{s}$),
  (4) BR($H^+ \to \tau^+ \nu_\tau$), (5) BR($H^+ \to W^+ h^0$), (6)
BR($H^+ \to W^+ A^0$).} \label{fig:br1a}
\end{figure}
\noindent {\bf Scenario A}. In Fig.~\ref{fig:br1a}(a) we present
the BRs for the channels $H^+ \to t \bar{b}$, $c \bar{b}$, $t
\bar{s}$, $ \tau^+ {\nu_\tau}$, $W^+ h^0$, $W^+ A^0$ as a function
of $m_{H^+}$, for $\tan\beta=0.1$ and fixing $m_{h^0}= 120$ GeV,
$m_{A^0}=300$ GeV and the mixing angle $\alpha = \pi /2$. When
$m_{H^+} < 175$ GeV , we can see that
the dominant decay of the charged Higgs boson is via the mode $c
\bar{b}$, with BR($H_i^+ \to c \bar{b}$) $\approx 1$, which will
have important consequences for charged Higgs boson production
through $c \bar{b}$-fusion at the LHC and may serve
as a distinctive feature of this model. For the case 175 GeV $< m_{H^+} <
180$ GeV the mode $t\bar{s}$ is relevant, which it is also very different
from the 2HDM-II and becomes an interesting phenomenological
consequence of the 2HDM-III. We can also observe that, for $m_{H^+}>
180$ GeV, the decay mode $ t \bar{b}$ is dominant (as in the 2HDM-II).
Now, from
Fig.~\ref{fig:br1a}(b), where $\tan \beta = 1$, we find that
 the dominant decay mode is into $\tau^+ {\nu_\tau}$ for the
range $m_{H^+}<$ 175 GeV, again for 175 GeV $< m_{H^+} < 180$ GeV
the mode $t\bar{s}$ is the leading one, but for 180
GeV$<m_{H^+}<600$ GeV, the decay channel $W^+ h^0$ becomes relevant,
whereas for the range 600 GeV $<m_{H^+}$ the mode $W^+ A^0$ is
dominant. It is convenient to mention that this sub-scenario is
special for the mode $t \bar{b}$, because its decay width is zero at
the tree-level, since the CKM contribution is canceled exactly with
the terms of the four-zero texture implemented for the Yukawa
coupling of the 2HDM-III. Then, see Fig.~\ref{fig:br1a}(c), for the
case with $\tan \beta = 15$ one gets that BR($H^+ \to
\tau^+{\nu_\tau}$) $\approx 1$ when $m_{H^+}< 180 $ GeV. However for
180 GeV$<m_{H^+}<300$ GeV, the dominant decay of the charged Higgs
boson is the mode  $t \bar{b}$, while in the range 300 GeV$<
m_{H^+}$, the decay channel $W^+ h^0$ is also relevant. For $\tan
\beta =70$, we show in plot Fig.~\ref{fig:br1a}(d) that the dominant
decay of the charged Higgs boson is the mode $\tau^+ {\nu_\tau}$,
when $m_{H^+}<300$ GeV, but that, for 300 GeV$< m_{H^+} <400$ GeV, the
decay channel $t \bar{b}$ becomes the leading one, whereas for the
range 400 GeV $ < m_{H^+}$, the mode $W^+ h^0$ is again dominant.

\begin{figure}
\centering
\includegraphics[width=6in]{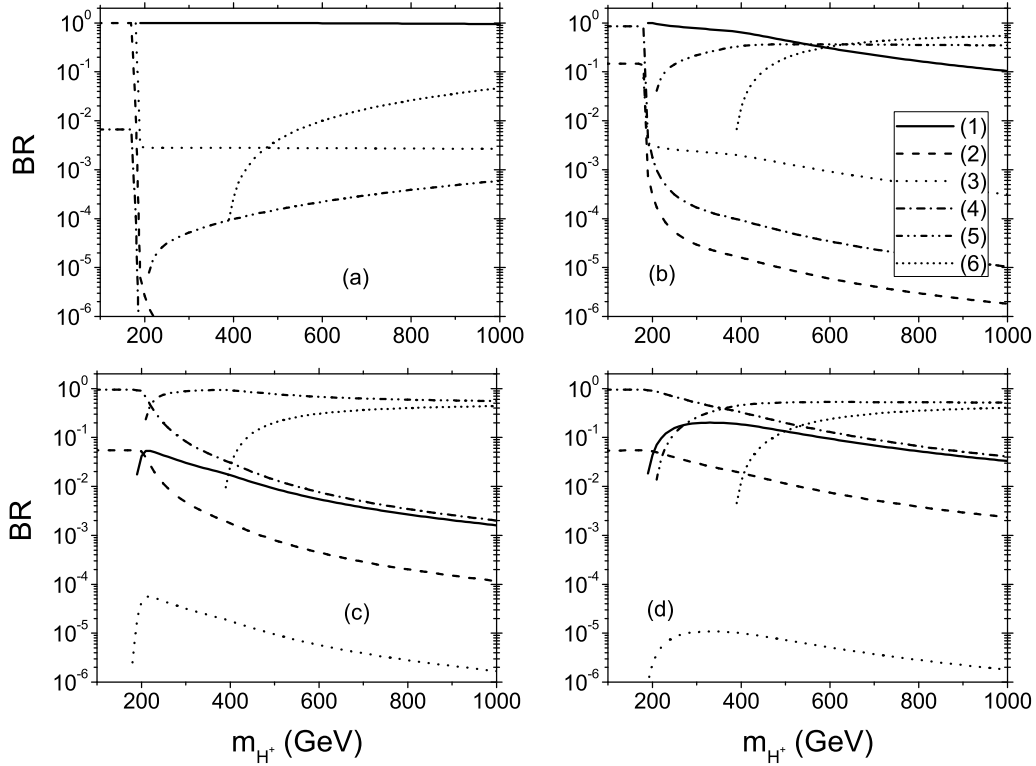}
\caption{The same as in Fig.~\ref{fig:br1a} but taking
$\tilde{\chi}_{ij}^{u}=0.1$, $\tilde{\chi}_{ij}^{d}=1$
(Scenario B).}
\label{fig:br1b}
\end{figure}
\noindent {\bf Scenario B}. In Fig.~\ref{fig:br1b}, we
present the BRs of the channels $H^+ \to t \bar{b}$, $c \bar{b}$,
$t \bar{s}$, $ \tau^+ {\nu_\tau}$, $W^+ h^0$, $W^+ A^0$ as a
function of $m_{H^+}$. From Fig.~\ref{fig:br1b}(a), we observe that
for $\tan \beta = 0.1$, when $m_{H^+} < 175$ GeV, the dominant decay
of the charged Higgs boson is the mode $c \bar{b}$, with BR($H_i^+
\to c \bar{b}$) $\approx 1$. When 175 GeV $< m_{H^+} < 180$ GeV the
mode $t\bar{s}$ is important and for $m_{H^+}> 180$ GeV the decay
mode $t \bar{b}$ becomes the leading one. From
Fig.~\ref{fig:br1b}(b), we see that, for $\tan \beta = 1$, the
dominant decay mode is now into $\tau^+ {\nu_\tau}$ for $m_{H^+}<$
175 GeV, while in the range 175 GeV $< m_{H^+} < 180$ GeV the
mode $t\bar{s}$ is relevant. For 180 GeV$<m_{H^+}<500$ GeV the decay
channel $t \bar{b}$ becomes the leading one, whereas for the range
500 GeV $<m_{H^+}$ the mode $W^+ A^0$ is dominant. From
Fig.~\ref{fig:br1b}(c), with $\tan \beta = 15$, one gets that
BR($H_i^+ \to \tau^+{\nu_\tau}$) $\approx 1$ for $m_{H^+}< 180 $
GeV. For 180 GeV$<m_{H^+}$, the dominant decay of  the
charged Higgs boson is instead the mode  $W^+ h^0$. Then, for $\tan \beta
=70$, we show in Fig.~\ref{fig:br1b}(d) that the dominant decay of
the charged Higgs state is via the mode $\tau^+ {\nu_\tau}$ when
$m_{H^+}<350$ GeV while for 350 GeV$< m_{H^+}$ the decay channel $W^+
h^0$ becomes the leading one.

\begin{figure}
\centering
\includegraphics[width=6in]{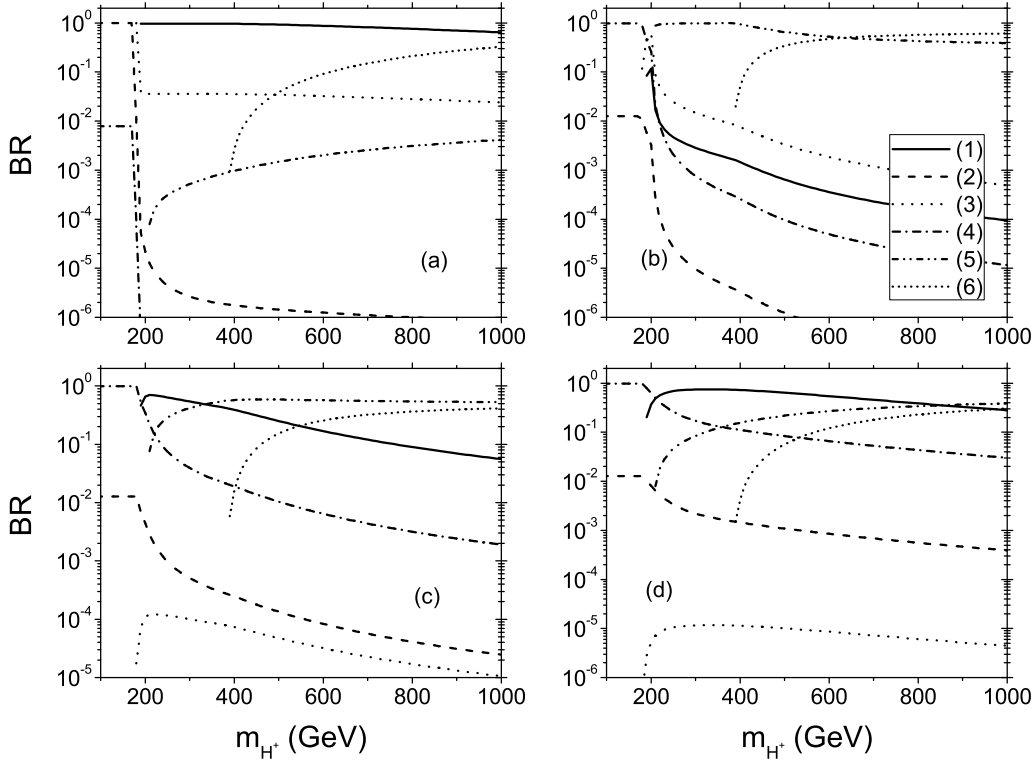}
\caption{The same as in Fig.~\ref{fig:br1a}, but taking
$\tilde{\chi}_{ij}^{u}=1$, $\tilde{\chi}_{ij}^{d}=0.1$
(Scenario C).}
\label{fig:br1c}
\end{figure}
\noindent {\bf Scenario C}. In Fig.~\ref{fig:br1c} we show the
corresponding plots for the BRs  of the  channels $H^+ \to t
\bar{b}$, $c \bar{b}$, $t \bar{s}$, $ \tau^+ {\nu_\tau}$, $W^+ h^0$,
$W^+ A^0$ as a function  of $m_{H^+}$. For $\tan \beta = 0.1$, as
one can see in Fig.~\ref{fig:br1c}(a), the mode $c\bar{b}$ is
dominant when $m_{H^+} < 170$ GeV, but for 175 GeV $< m_{H^+} < 180$
GeV the mode $t\bar{s}$ is relevant, while for 180 GeV$< m_{H^+}$
the mode $t\bar{b}$ becomes dominant. For $\tan \beta =1$, we
observe from Fig.~\ref{fig:br1c}(b) that the dominant decay modes
are: $ \tau^+ {\nu_\tau}$ in the range $m_{H^+} < 170$ GeV,
$t\bar{s}$ for 175 GeV $< m_{H^+} < 180$ GeV, $W^+ h^0$ for 180 GeV
$< m_{H^+} < 600$ GeV and $W^+ A^0$ when 600 GeV $< m_{H^+}$. For
$\tan \beta = 15$, as shown in Fig.~\ref{fig:br1c}(c), the relevant
decay channels are: $ \tau^+ {\nu_\tau}$ in the range $m_{H^+} <
180$ GeV, $t \bar{b}$ when 180 GeV $<m_{H^+}<300$ GeV, $W^+ h^0 $
for 300 GeV $<m_{H^+}$. In Fig.~\ref{fig:br1c}(d), for $\tan \beta
=70$, we observe that  $ \tau^+ {\nu_\tau}$ dominates
when $m_{H^+} < 180$ GeV, but when 180 GeV
$<m_{H^+}<900$ GeV the mode $t\bar{b}$ is the leading one, whereas
for $900$  GeV $<m_{H^+} $ the mode $W^+ h^0$ is the most relevant one.

\begin{figure}
\centering
\includegraphics[width=6in]{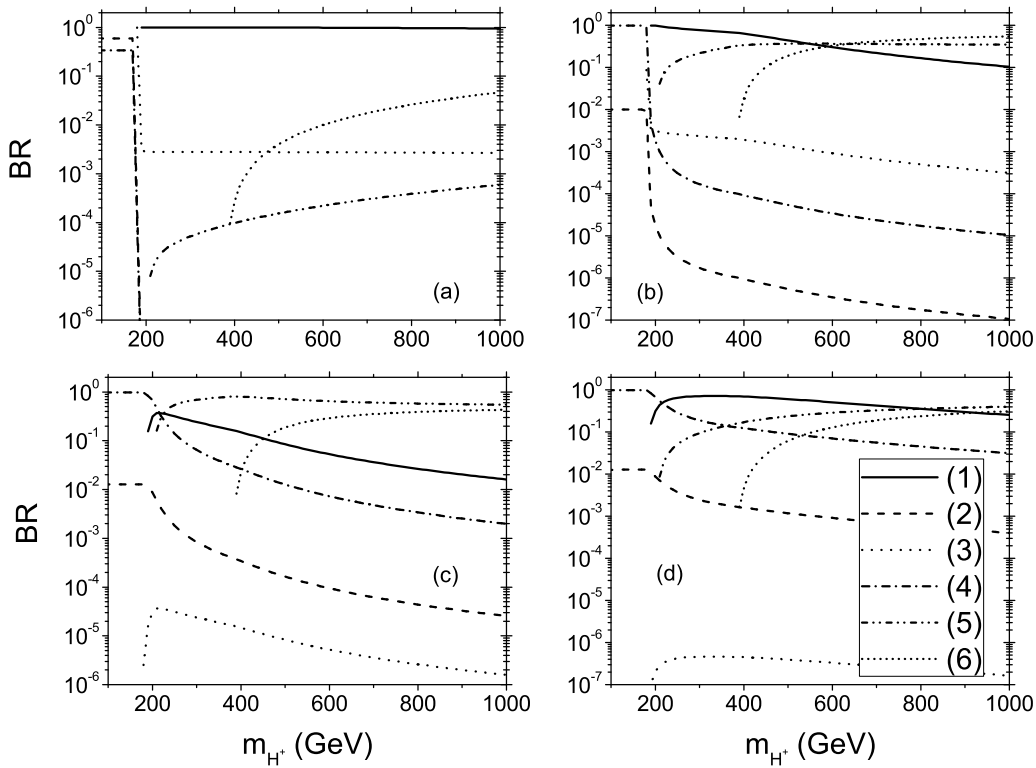}
\caption{The same as in Fig.~\ref{fig:br1a}, but taking
$\tilde{\chi}_{ij}^{u}=0.1$, $\tilde{\chi}_{ij}^{d}=0.1$
(Scenario D).}
\label{fig:br1d}
\end{figure}
\noindent {\bf Scenario D}. In  Fig.~\ref{fig:br1d} we present
plots for the BRs  of the channels $ t \bar{b}$, $c \bar{b}$, $t
\bar{s}$, $ \tau^+ {\nu_\tau}$, $W^+ h^0$, $W^+ A^0$ as a function
of $m_{H^+}$. For $\tan \beta = 0.1$, we show in
Fig.~\ref{fig:br1d}(a) that the dominant decay modes for the $H^+$ are:
$c\bar{b}$ in the range $m_{H^+}<175 $ GeV, $t\bar{s}$ when 175 GeV
$<m_{H^+}< 180$ GeV, $t\bar{b}$ for $ 180$ GeV $<m_{H^+}$. For $\tan
\beta = 1$, we show in Fig.~\ref{fig:br1d}(b) that the mode $ \tau^+
{\nu_\tau}$ is dominant in the range $m_{H^+} < 175$ GeV, whereas
for  175 GeV $<m_{H^+}< 180$ GeV the relevant decay channel is
$t\bar{s}$, whilst the mode $ t \bar{b}$ dominates for 180 GeV $<
m_{H^+} < 550$ GeV and the mode $W^+ A^0$ does so when 550 GeV $<
m_{H^+}$. For $\tan \beta = 15$, we observe in Fig.~\ref{fig:br1d}(c)
that the relevant decay channels are: $ \tau^+
{\nu_\tau}$ in the range $m_{H^+} < 250$ GeV and $W^+ h^0 $ for 250
GeV $<m_{H^+}$. Finally, for $\tan \beta =70$, see
Fig.~\ref{fig:br1d}(d), we obtain that, when $m_{H^+} < 230$ GeV,
the mode $ \tau^+ {\nu_\tau}$ becomes the most important one but, for
230 GeV $<m_{H^+}<800$ GeV, the channel $t\bar{b}$ is the leading one,
whereas,
for 800 GeV  $<m_{H^+}$, the mode $W^+ h^0$ is the dominant one.\\
\begin{figure}
\centering
\includegraphics[width=6in]{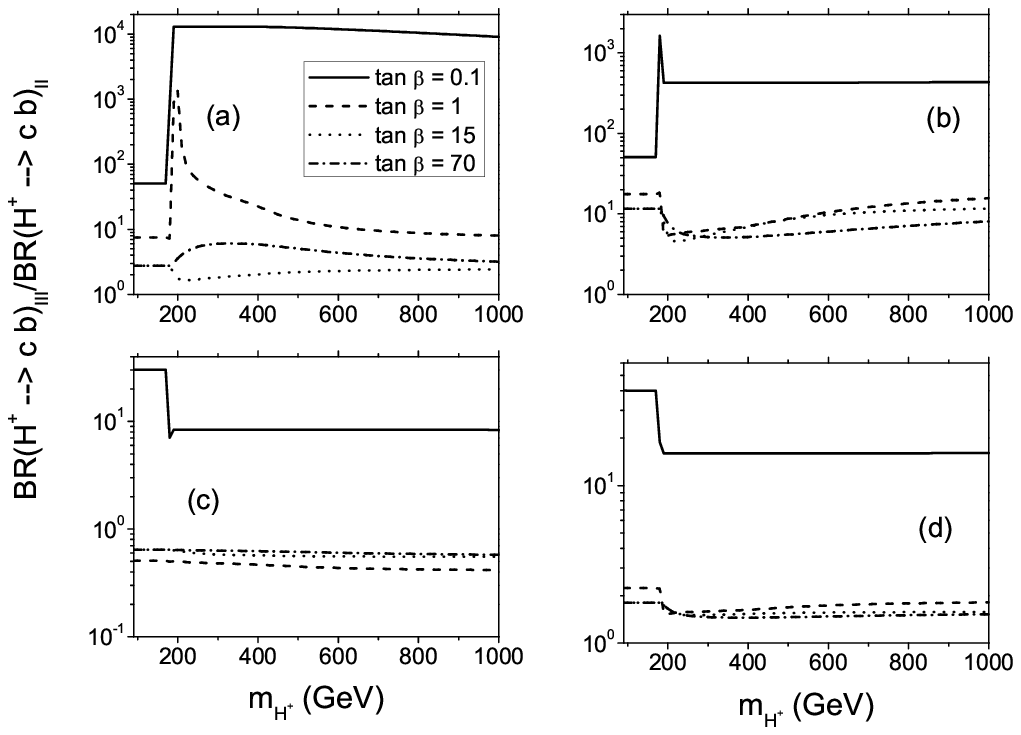}
\caption{The figure shows the BR$(H^+ \to c \bar{b})_{\rm III}/{\rm BR} (H^+ \to
c \bar{b})_{\rm II} $ {\it vs.} $m_{H^+}$, taking $\tan \beta =0.1, \,
1, \, 15, \, 70 $ for (a)$\tilde{\chi}_{ij}^{u,d}=1$, (b)
$\tilde{\chi}_{ij}^{u,d}=-1$, (c)$\tilde{\chi}_{ij}^{u,d}=0.1$, (d)
$\tilde{\chi}_{ij}^{u,d}=-0.1$.} \label{fig:Rcb1}
\end{figure}
\begin{figure}
\centering
\includegraphics[width=6in]{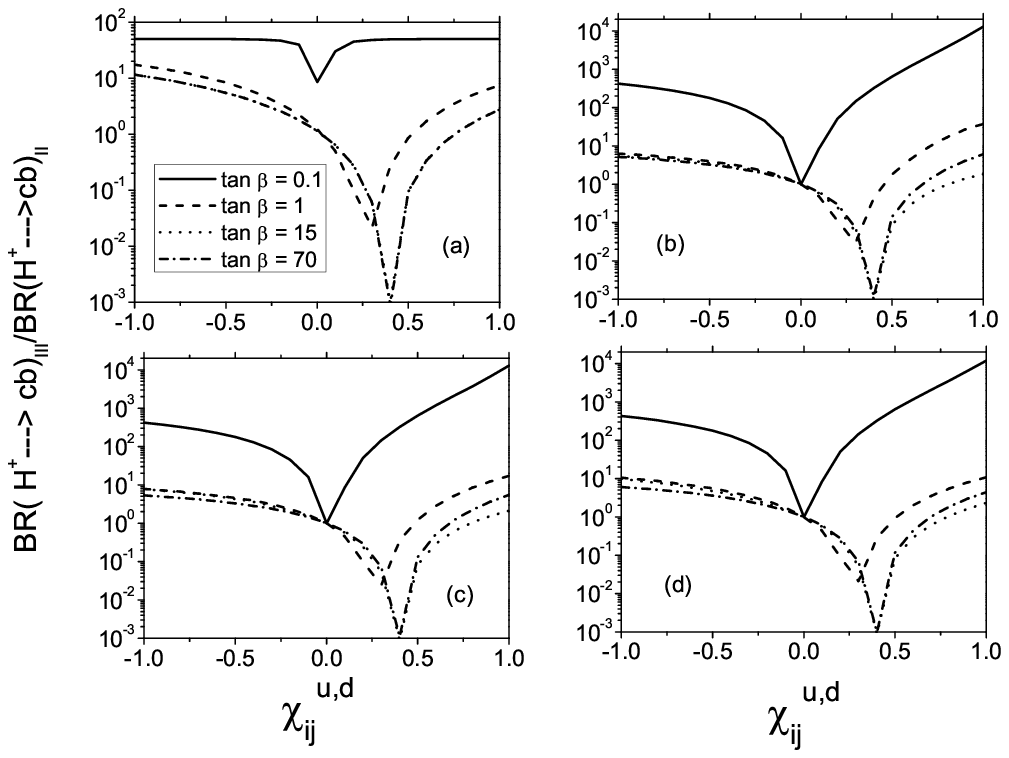}
\caption{The figure shows the BR$(H^+ \to c \bar{b})_{\rm III}/{\rm BR} (H^+ \to
c \bar{b})_{\rm II} $ {\it vs.} $\tilde{\chi}_{ij}^{u,d}$, taking $\tan
\beta =0.1, \, 1, \, 15, \, 70 $ for (a)$m_{H^+}=150$ GeV, (b)
$m_{H^+}=300$ GeV, (c)$m_{H^+}=450$ GeV, (d) $m_{H^+}=600$ GeV.}
\label{fig:Rcb2}
\end{figure}
\vspace*{0.15cm}

In order to cover further the Higgs sector in our
analysis, it is appropriate to also mention how the previous results
change with $m_{h^0}$, $m_{A^0}$ and $\alpha$. Regarding the former two,
clearly, the lather the neutral Higgs boson mass the later the corresponding
$H^\pm$ decay channel will onset. Regarding the latter, we adopted two further
choices,  $\alpha = \beta$ and $0$, in all scenarios
previously studied. In general the behavior of the decay modes of the
charged Higgs boson is similar to the cases presented above, except
for the decay channel $Wh^0$. For $\alpha = 0$, this mode has
BR $< 10 ^ {-3} $ when $\tan \beta$ is large. However, for $\tan
\beta <1 $, it becomes the dominant one. In the case $\alpha = \beta$,
the decay channel $Wh^0$ can be dominant with a BR that could be $O(1)$. \\

\begin{figure}
\centering
\includegraphics[width=6in]{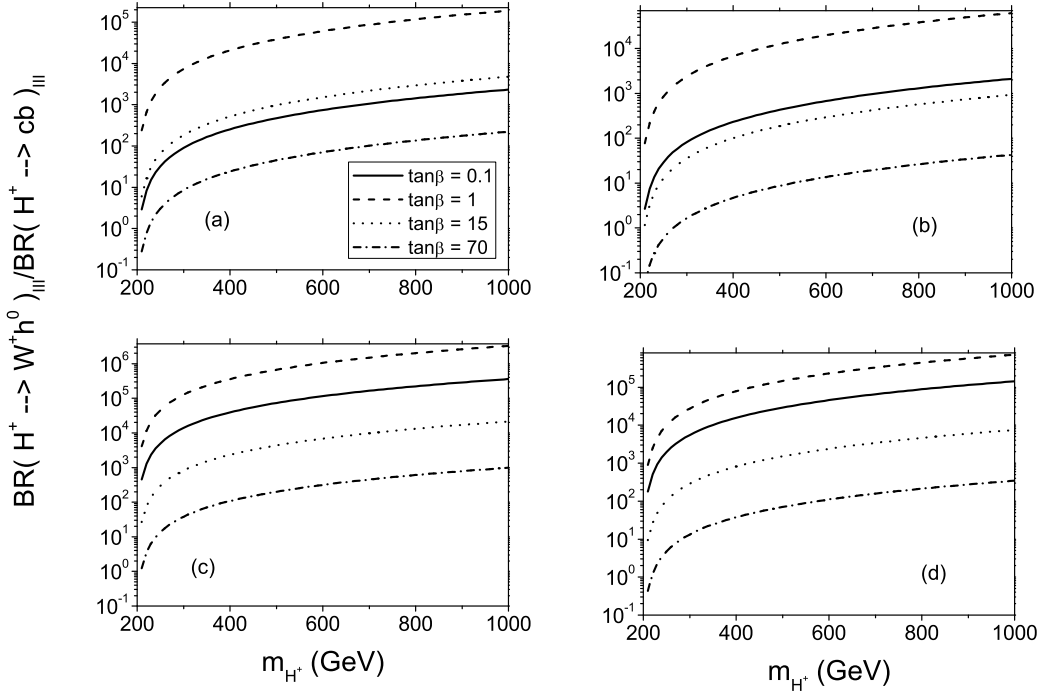}
\caption{The figure shows the BR$(H^+ \to W^+ h^0)_{\rm III}/{\rm BR} (H^+ \to c
\bar{b})_{\rm III} $ {\it vs.} $m_{H^+}$, taking $\tan \beta =0.1, \, 1,
\, 15, \, 70 $ for (a)$\tilde{\chi}_{ij}^{u,d}=0.1$, (b)
$\tilde{\chi}_{ij}^{u,d}=-0.1$, (c)$\tilde{\chi}_{ij}^{u,d}=1$, (d)
$\tilde{\chi}_{ij}^{u,d}=-1$.} \label{fig:Rcbwh1}
\end{figure}

As a general lesson from this section, and distinctive features of
our 2HDM-III, we can see that both decay modes  $W^+ h^0$ and $c
\bar{b}$ become very relevant phenomenologically, effectively of
$O(1)$ for some of the scenarios considered. Therefore, we want to
study next the general behaviour of these decay modes, in relation
to the 2HDM-II case. In order to compare the 2HDM-III results with
those in the 2HDM-II, we show in Fig.~\ref{fig:Rcb1} the ratio
BR$(H^+ \to c \bar{b})_{\rm III}/{\rm BR} (H^+ \to c \bar{b})_{\rm
II} $ {\it vs.} $m_{H^+}$, taking again $\tan \beta =$ 0.1, 1, 15,
70, for: (a)$\tilde{\chi}_{ij}^{u,d}=1$, (b)
$\tilde{\chi}_{ij}^{u,d}=-1$, (c) $\tilde{\chi}_{ij}^{u,d}=0.1$
and (d) $\tilde{\chi}_{ij}^{u,d}=-0.1$. We observe that the mode
$c \bar{b}$ is important when 200 GeV $<m_{H^+} <$ 300 GeV and for
$0.1 \leq \tan \beta \leq 1$, taking $\tilde{\chi}_{ij}^{u,d}=1$.
Now, in Fig.~\ref{fig:Rcb2}, we present the behaviour of the ratio
BR$(H^+ \to c \bar{b})_{\rm III}/{\rm BR} (H^+ \to c \bar{b})_{\rm
II} $ as a function of $\chi_{i,j}^{u,d}$, for the cases: (a)
$m_{H^+}=150$ GeV, (b) $m_{H^+}=300$ GeV, (c) $m_{H^+}=450$ GeV
and (d) $m_{H^+}=600$ GeV. Again, one can see that the largest
enhancement arises when $m_{H^+}=300$ GeV and
$\tilde{\chi}_{ij}^{u,d}=1$.  Finally, specific to the 2HDM-III,
we show in Fig.~\ref{fig:Rcbwh1} the ratio BR$(H^+ \to W^+
h^0)_{\rm III}/{\rm BR} (H^+ \to c \bar{b})_{\rm III} $ {\it vs.}
$m_{H^+}$, taking $\tan \beta =$ 0.1, 1, 15, 70, for:
(a)$\tilde{\chi}_{ij}^{u,d}=0.1$, (b)
$\tilde{\chi}_{ij}^{u,d}=-0.1$, (c) $\tilde{\chi}_{ij}^{u,d}=1$
and (d) $\tilde{\chi}_{ij}^{u,d}=1$. We find that BR$(H^+ \to W^+
h^0)_{\rm III}$ is much larger than BR$(H^+ \to c \bar{b})_{\rm
III}$ when $\tilde{\chi}_{ij}^{u,d}=1$ and the mass of the charged
Higgs boson is close to the upper limit obtained by unitarity
conditions, which is about 800 GeV \cite{unitarity}. Thus, we find
that the effect of the modified Higgs couplings typical of the
2HDM-III
 shows up clearly in the pattern of charged Higgs
boson decays, which can be very different from the 2HDM-II case and thus
 enrich the possibilities to search for
$H^{\pm}$ states at current (Tevatron) and future (LHC, ILC/CLIC) machines.
\section{\bf Charged Higgs boson production at LHC }

The production of charged Higgs bosons at hadron colliders  has been evaluated in
early \cite{ldcysampay} (also for the Superconducting Super Collider, SSC)
and more recent \cite{newhcprod} (for the LHC) literature, mainly for the 2HDM-II
and its SUSY realization (i.e., the MSSM). In these two scenarios,
when kinematically allowed, the top quark decay
channel $t \to b H^+$ is the dominant $H^\pm$ production mechanism.
Instead, above the threshold for such a decay, the dominant $H^\pm$ production reaction is
gluon-gluon fusion into a 3-body final state, {i.e.}, $gg \to tbH^\pm$\footnote{In fact,
these two mechanisms are intimately related, see below.}. Both processes depend
on the coupling $H^- t\bar b$ and  are therefore sensitive to the
modifications that arise in the 2HDM-III for this vertex. However,
detection of the final state will depend on the charged Higgs boson
decay mode, which could include a complicated final state, that
could in turn be difficult to reconstruct. For these reasons, it is very
important to look for other production channels, which may be easier
to reconstruct. In this regard, the $s$-channel production of charged
Higgs bosons, through the mechanism of $c\bar b$-fusion,  could help to make
more viable the detection of several charged Higgs boson decay
channels \cite{He:1998ie}.

Here, we shall evaluate the predictions of the 2HDM-III for the $t
\to b H^+$  (and $s H^+$) decay rate plus the $c\bar{b}$- as well
as the $gg$-fusion mechanisms (hereafter, referred to as `direct'
and `indirect' $H^\pm$ production, respectively).

\subsection{The decays $t \to H^+ \, b$, $H^+ s$}
We shall  discuss here the charged Higgs boson interactions with
heavy quarks ($t,b,c,s$) and their implications for charged Higgs boson
production through top quark decays. In order to study the top quark
BRs, besides the SM decay mode $t \to  b W^+$, we need to
consider both decays $t \to b H^+$ and $t \to s H^+$, because
these modes could both be important for several parameter configurations
within our model.  The decay width
of these modes takes the following form:
\begin{eqnarray}
\Gamma(t \to  d_jH^+) &=& \frac{g^2}{128 \pi m_W^2 m_t^3 } \lambda^{1/2}(m_t^2,m_{H^+}^2,m_b^2) \nonumber\\
& & \times \, \bigg( \bigg[ (m_t+m_b)^2-m_{H^+}^2\bigg] S_{3j}^2+
\bigg[ (m_t+m_b)^2-m_{H^+}^2\bigg] P_{3j}^2 \bigg),
\end{eqnarray}
where $\lambda$ is the usual kinematic factor $\lambda(a,b,c)=
(a-b-c)^2-4bc$, $j=2$ for the mode $sH^+$ and $j=3$ for the mode
$bH^+$. Furthermore, we shall neglect the decay width for the light
fermion generations.  If one takes $\tilde{\chi}_{i,j} \to 0$,
the formulae for the decay width reduce to the 2HDM-II case:
see, e.g., \cite{kanehunt}.

We have explored several theoretically allowed regions within our
scenario, which are constrained by using the bounds on the
BR$(t \to  b H^+ )$. In the so-called ``tauonic Higgs model''
\cite{Abulencia:2005jd}, the decay mode ($H^+ \to \tau^+ \,
{\nu}_{\tau}$) dominates the charged Higgs boson decay width, and
BR$(t \to b H^+)$ is constrained to be less than 0.4 at 95 \%
C.L. \cite{Abulencia:2005jd}. However, if no assumption is made on
the charged Higgs boson decay, BR$(t \to b H^+)$ is constrained
to be less than 0.91 at 95 \% C.L. \cite{Abulencia:2005jd}. However,
the combined LEP data exclude a charged Higgs boson
with mass less than 79.3 GeV at 95 \% C.L., a limit valid for an
arbitrary BR$(H^+ \to \tau^+ \, {\nu_\tau})$ \cite{partdat}.
Thus, in order to perform our analysis, we need to discuss all the
charged Higgs boson decays following the steps of our previous
paper \cite{ourtriplets}. In the present section, we take
 all charged Higgs boson decays relevant for masses below that
of the top quark, thus including the modes $ \tau^+ {\nu_\tau}, t
\bar{s},  c \bar{b}, W^+ h^0, W^+ A^0$.  As usual, we refer to our
four benchmark scenarios.

\begin{figure}
\centering
\includegraphics[width=6in]{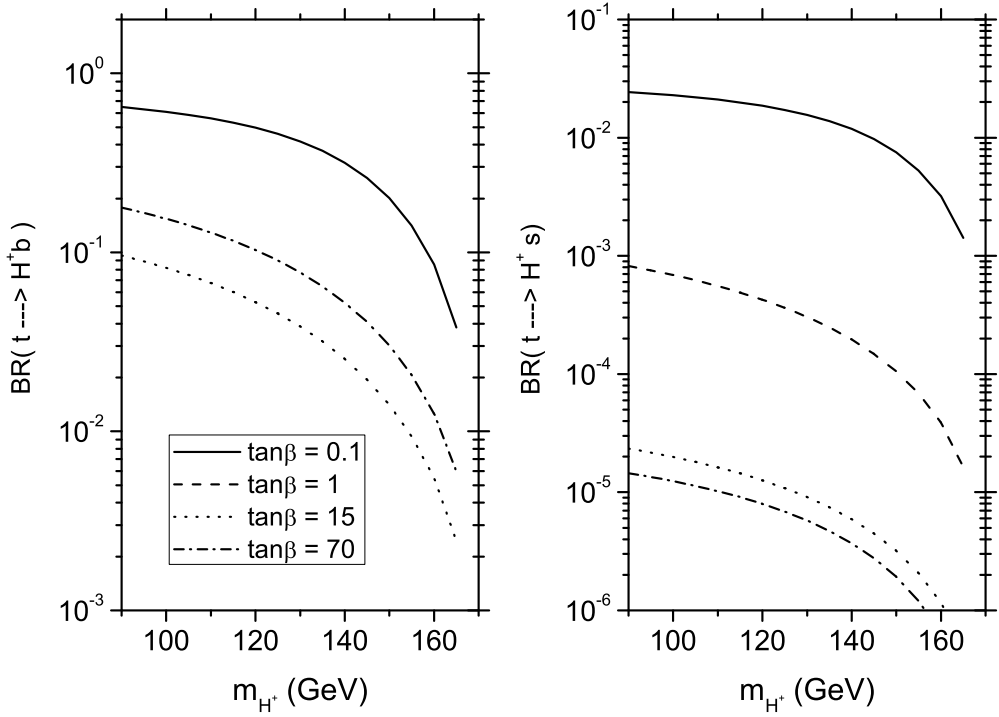}
\caption{ It is plotted: the BR($t \to b \, H^+ $) {\it vs.}
$m_{H^+}$ (left); the BR($t \to s \, H^+ $) {\it vs.} $m_{H^+}$
(right). Here is for Scenario A, obtained by taking
$\tilde{\chi}_{ij}^u = 1$ and $\tilde{\chi}_{ij}^d = 1$, for $\tan
\beta=0.1$  (solid), $1$ (dashes), $15$ (dots) and $70$
(dashes-dots).} \label{fig:brt1a}
 \end{figure}

\noindent {\bf Scenario A.} Remember that this scenario was defined
by taking $\tilde{\chi}_{ij}^u=1$ and $\tilde{\chi}_{ij}^d=1$, while
for $\tan \beta $ we considered the values $\tan \beta=0.1,\,1,\,15,
\, 70$. In Fig.~\ref{fig:brt1a} we present plots of BR($t \to
b \, H^+$) {\it vs.} $m_H^+$ and BR($t \to s \, H^+ $) {\it vs.}
$m_{H^+}$. We can observe that a charged Higgs boson within the mass
range 80 GeV $< m_{H^+} < 170$ GeV and for $1 < \tan\beta < 70$
satisfies the constraint BR$(t \to b \, H^+) < 0.4$. Furthermore,
from the plots of Fig.~\ref{fig:br1a}, we can see that in this
scenario the dominant decay mode is into $\tau^+{\nu}_\tau$ for
$\tan \beta =1$, 15, 70, therefore we fall within the scope of the
tauonic Higgs model, so that BR$(t \to H^+ \, b) \leq 0.4$ applies.
However, for the case $\tan \beta = 0.1$, the dominant decay of the
charged Higgs boson is the mode $c \bar{b}$ and the mode $t \to b
\, H^+$ satisfies the constraint BR$(t \to b \, H^+) < 0.9$ in the
range described above.

\begin{figure}
\centering
\includegraphics[width=5in]{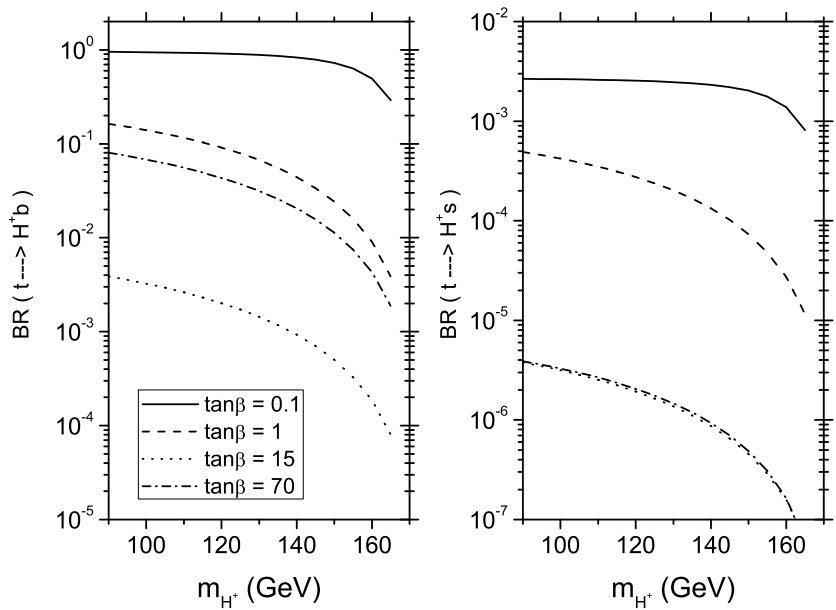}
\caption{The same as in Fig.~\ref{fig:brt1a} but taking
$\tilde{\chi}_{ij}^u = 0.1$ and $\tilde{\chi}_{ij}^d = 1$ (Scenario B).}
\label{fig:brt1b}
\end{figure}

\noindent {\bf Scenario B.} In Fig.~\ref{fig:brt1b} we present
similar plots for the case $\tilde{\chi}_{ij}^u=0.1$ and
$\tilde{\chi}_{ij}^d=1$, taking $\tan \beta=0.1,\,1,\,15, \, 70$. We
can observe that the mode $t \to b \, H^+$ satisfies the constraint
BR$(t \to b \, H^+) < 0.4$ within the ranges 80 GeV $< m_{H^+} <
170$ GeV and $1 < \tan\beta < 70$. Thus, from Fig.~\ref{fig:br1b},
we can see that in this range the dominant decay mode is into
$\tau^+{\nu}_\tau$, therefore this setups also falls within the realm of the
tauonic Higgs model, so that BR$(t \to H^+ \, b) \leq 0.4$ must hold
in this scenario. For $\tan \beta = 0.1$, the dominant decay of the
charged Higgs boson is $c \bar{b}$, thus the channel $t \to b
\, H^+$ must satisfy the constraint BR$(t \to b \, H^+) < 0.9$,
which is fulfilled in the range studied.

\begin{figure}
\centering
\includegraphics[width=5in]{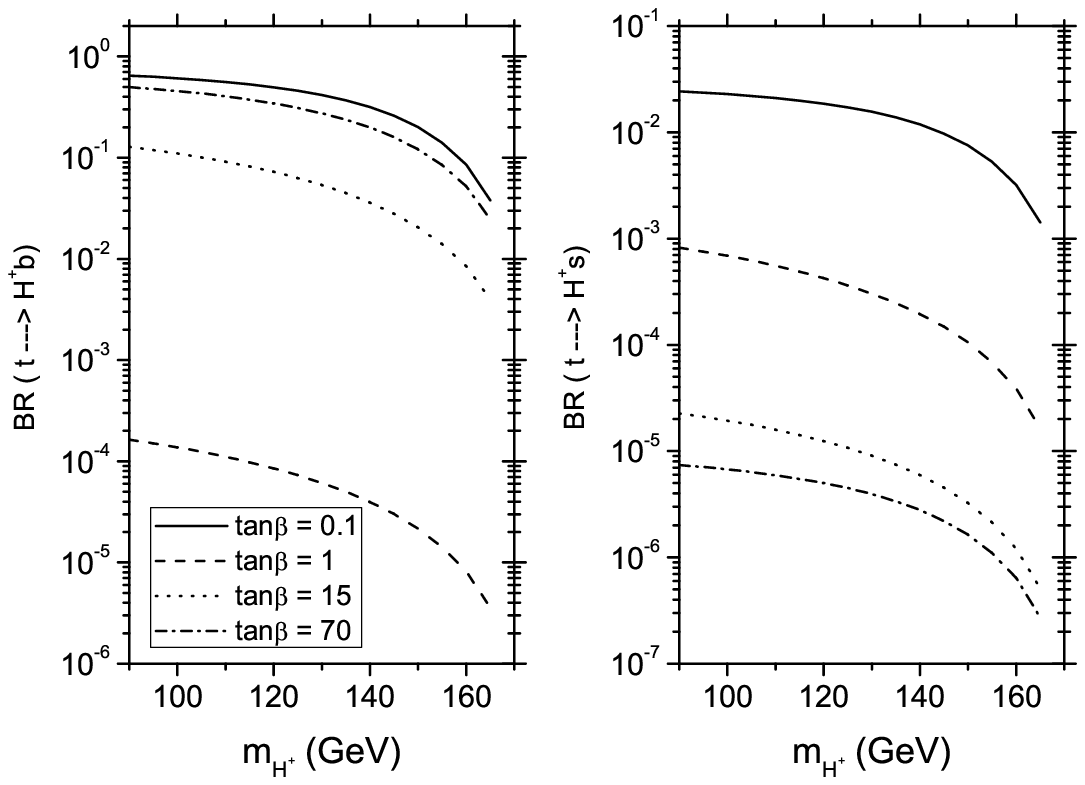}
\caption{The same as in Fig.~\ref{fig:brt1a} but taking
$\tilde{\chi}_{ij}^u = 1$ and $\tilde{\chi}_{ij}^d =0.1$  (Scenario C).}
\label{fig:brt1c}
\end{figure}

\noindent {\bf Scenario C.} In Fig.~\ref{fig:brt1c} we present the
corresponding plots for the case $\tilde{\chi}_{ij}^u=1$ and
$\tilde{\chi}_{ij}^d=0.1$, taking again $\tan \beta=0.1,\,1,\,15,
\, 70$. We can observe that the mode $t \to b \, H^+$ satisfies
the constraint BR$(t \to b \, H^+) < 0.4$ in the range 80 GeV $<
m_{H^+} < 170$ GeV and $1 < \tan\beta < 70$. Similarly, as in
scenario A, for $\tan \beta = 0.1$ the dominant decay of the
charged Higgs boson is the mode  $c \bar{b}$, thus the mode $t \to
b \, H^+$ must satisfies the constraint BR$(t \to b \, H^+) <
0.9$, indeed satisfied in the range analyzed here.

\begin{figure}
\centering
\includegraphics[width=5in]{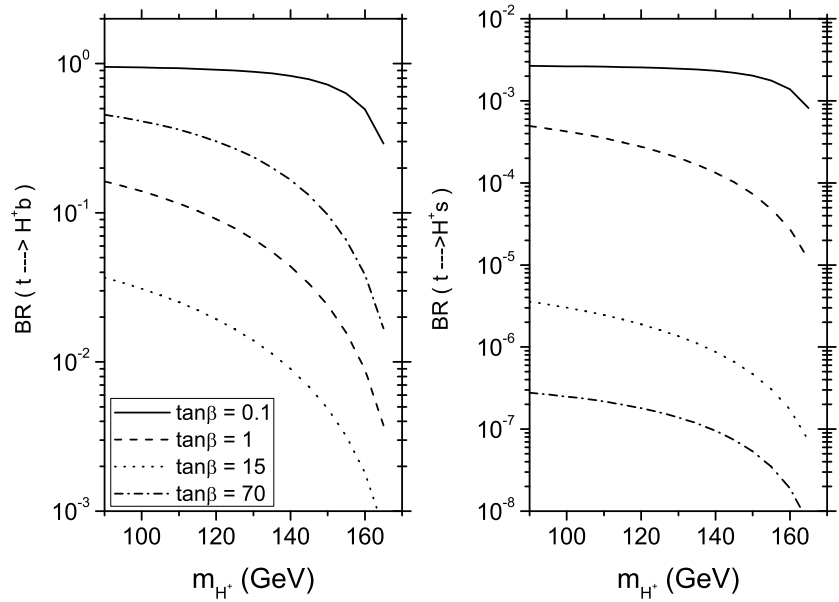}
\caption{The same as in Fig.~\ref{fig:brt1a} but taking
$\tilde{\chi}_{ij}^u = 0.1$ and $\tilde{\chi}_{ij}^d = 0.1$  (Scenario D).}
\label{fig:brt1d}
\end{figure}

\noindent{\bf Scenario D.} Recall that this was defined by taking
$\tilde{\chi}_{ij}^u=0.1$ and $\tilde{\chi}_{ij}^d=0.1$. In
Fig.~\ref{fig:brt1d} we present the usual plots of the BR($t \to b \,
H^+ $) and BR$(t\to bH^+)$ {\it vs.} $m_{H^+}$. One can see that, for charged Higgs
boson masses within the range 80 GeV $< m_{H^+} < 170$ GeV and $1 <
\tan\beta < 70$, the model fulfills the constraint BR$(t \to b \,
H^+) < 0.4$. Furthermore, for $\tan \beta = 0.1$, the dominant decay of the
charged Higgs boson is the mode  $c \bar{b}$, thus the mode $t \to b
\, H^+$ satisfies the constraint BR$(t \to b \, H^+) < 0.9$ in the
range studied.

In short, as bottomline of these execises,
we have identified regions of the 2HDM-III parameter space where a charged
Higgs mass below $m_t-m_b$ has not been excluded by Tevatron.
Therefore, the LHC is best positioned in order to probe charged
Higgs bosons with such masses.

\subsection{\bf Direct production of charged Higgs bosons at the LHC}

\begin{figure}
\centering
\includegraphics[width=5in]{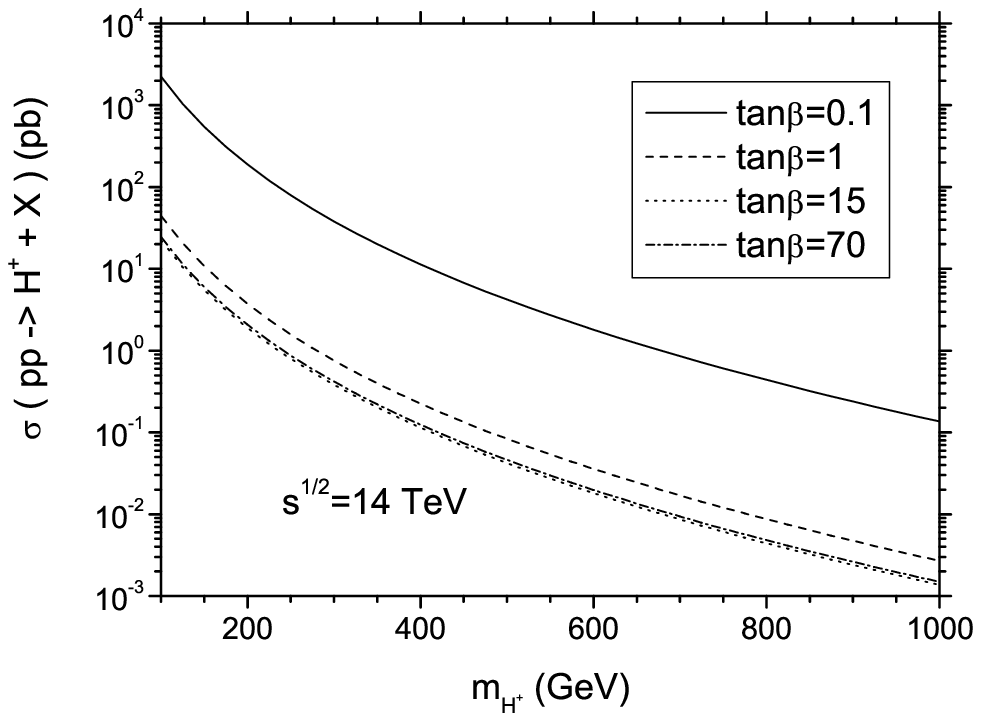}
\caption{ The figure shows the total cross section rates of process
$h_1 h_2 (c\bar{b}) \to H^+ X$ as a function of $m_{H^+}$ in the
2HDM-III at LHC energies ($\sqrt s=14$ TeV), by taking
$\tilde{\chi}^d_{l3}=1$ and $\tilde{\chi}^u_{2l}=1$ ($l=1,2,3$). The
lines correspond to: $\tan\beta=0.1$, $\tan\beta=1$, $\tan\beta=15$, $\tan\beta=70$. } \label{fig:1}
\end{figure}

\begin{figure}
\centering
\includegraphics[width=5in]{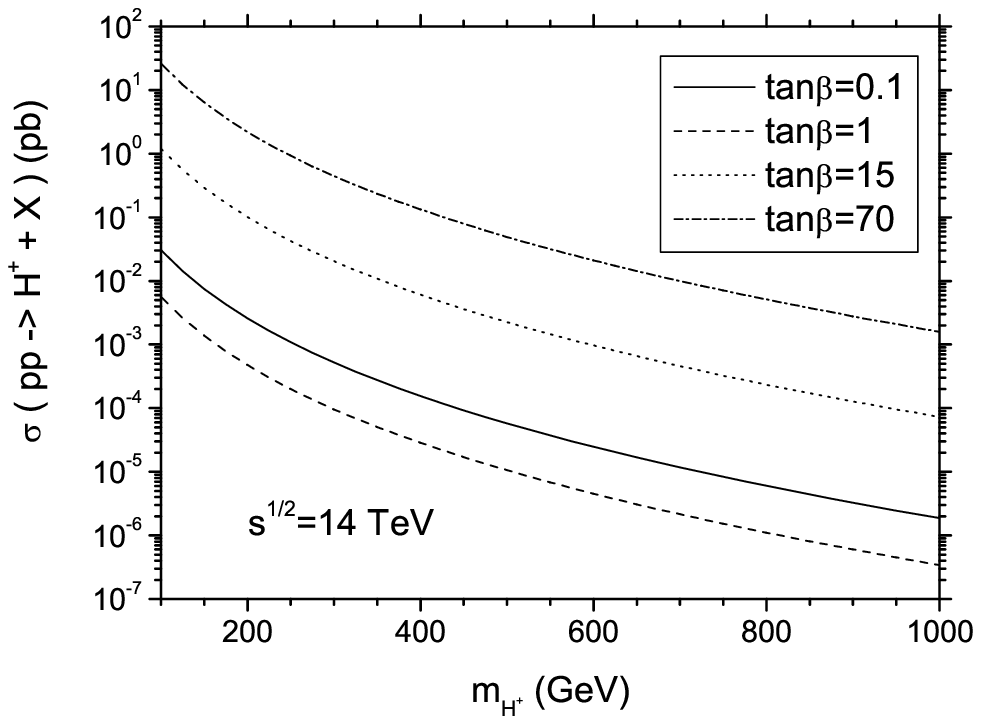}
\caption{ The figure shows the total cross section rates of process
$h_1 h_2 (c\bar{b}) \to H^+ X$ as a function of $m_{H^+}$ in the
2HDM-II at LHC energies ($\sqrt s=14$ TeV), by taking $(V_{\rm CKM})_{23}=4.16
\times 10^{-2}$ and $(V_{\rm CKM})_{33} \approx 1$. The lines correspond
to: $\tan\beta=0.1$, $\tan\beta=1$, $\tan\beta=15$,
$\tan\beta=70$. } \label{fig:2}
\end{figure}

The $H^{\pm} \bar{q} q^{\prime}$ vertex with large flavor mixing coupling,
that arises in the 2HDM-III, enables the possibility of studying the
production of charged Higgs boson via the $s$-channel production
mechanism, $c\bar{b} \to H^{+}$ + c.c. This process was
discussed first by Ref.~\cite{He:1998ie}, both within
topcolor models and a simplified version of the 2HDM-III. Then the
SUSY case was discussed in \cite{DiazCruz:2001gf} and
\cite{Dittmaier:2007uw}. Here we perform a detailed study of this
mechanism within the 2HDM-III, paying special attention to the
effects induced by the assumed Yukawa texture on the charged Higgs boson
couplings. Defining the $H^{\pm} \bar{q} q^{\prime}$ coupling here
as $C_L \frac{1- \gamma_{5}}{2} + C_R \frac{1+ \gamma_{5}}{2}$, we
can express the total cross section for $H^+$ direct production at hadron
colliders as \cite{He:1998ie}

\begin{equation}
\label{sigma} \sigma(h_1 h_2 (c\bar{b}) \to H^+ X) =
\frac{\pi}{12s}(|C_L|^2+|C_R|^2) \, I^{h_1,h_2}_{c,\bar{b}},
\end{equation}
\noindent where
\begin{equation}
\label{sigma2} I^{h_1,h_2}_{c,\bar{b}} = \int_{\tau}^{1}
\frac{dx}{x} \, [f^{h_1}_c(x,\tilde{Q}^2)
f^{h_2}_{\bar{b}}(\tau/x,\tilde{Q}^2)+f^{h_1}_{\bar{b}}(x,\tilde{Q}^2)
f^{h_2}_c(\tau/x,\tilde{Q}^2)]
\end{equation}
\noindent and $\tau=m_{H^{\pm}}^2/s$. The Parton Distribution
Functions (PDFs) $f^{h_i}_{q}(x,\tilde{Q}^2)$ used here are from
\cite{pumplin}, with scale choice $\tilde{Q}^2=m_{H^{+}}^2$.

From Eq.~(\ref{LCCH}) we see that, for the case of the
2HDM-III, $C_L$ and $C_R$ entering the subprocess $c\bar{b} \to H^+$
are given by

\begin{equation}
\label{CLfactor} C_L\equiv C^{\rm III}_L  = - \frac{ig}{\sqrt{2}
M_W} \sum_{l=1}^{3} \left[ \cot \beta \, m_{c} \, \delta_{2l}
  -\frac{\csc \beta}{\sqrt{2} }  \,\sqrt{m_{c} m_{u_l} } \, \tilde{\chi}^u_{2l} \right]
  (V_{\rm CKM})_{l3} , \\
\end{equation}
\noindent and
\begin{equation}
\label{CRfactor} C_R\equiv C^{\rm III}_R  = - \frac{ig}{ \sqrt{2}
M_W} \sum_{l=1}^{3}
 \left[ \tan \beta \, m_{d_{l}} \, \delta_{l3}
-\frac{\sec \beta}{\sqrt{2} }  \,\sqrt{m_{d_l} m_{d_3} } \,
\tilde{\chi}^d_{l3}
\right] (V_{\rm CKM})_{2l} . \\
\end{equation}

We notice here that Eqs. (\ref{CLfactor}) and (\ref{CRfactor}) reduce
to the case of the 2HDM-II if one takes

\begin{equation}
\label{CLfactor2} C_L\equiv C^{\rm II}_L  = - \frac{ig}{\sqrt{2} M_W} \cot \beta
\,
m_{c} \, (V_{\rm CKM})_{23} \\
\end{equation}
\noindent and
\begin{equation}
\label{CRfactor2} C_R\equiv C^{\rm II}_R  = - \frac{ig}{ \sqrt{2} M_W}
 \tan \beta \, m_{b}(V_{\rm CKM})_{23}. \\
\end{equation}

In Fig.~\ref{fig:1}, we present plots for the total cross
section rates of process $h_1 h_2 (c\bar{b}) \to H^+ X$ as a
function of $m_{H^+}$ in the framework of the  2HDM-III, by taking
$\tilde{\chi}^d_{l3}=1$ and $\tilde{\chi}^u_{2l}=1$ ($l=1,2,3$),
at LHC energies ($\sqrt s=14$ TeV), for the cases: (a) $\tan\beta=0.1$,
(b) $\tan\beta=1$, (c) $\tan\beta=15$, (d) $\tan\beta=70$. The
sum over $l$ is performed over all the three quark families and we
take for the quark masses: $m_u=2.55$ MeV, $m_d=5.04$ MeV,
$m_c=1.27$ GeV, $m_s=104$ MeV, $m_b=4.20$ GeV, and
$m_t=171.2$ GeV \cite{partdat}. We have checked numerically that
the term proportional to $\frac{1}{2} \,\csc^2 \beta \, m_{c}
m_{t} \, |\tilde{\chi}^u_{23} \, (V_{\rm CKM})_{33}|^2$ provides the
most important contribution to the cross section rates and
dominates by far for $\tilde{\chi}^u_{23} \approx 1$. On the other
hand, the expected integrated luminosity at LHC is of the order
$10^5$ pb$^{-1}$ and given that $\sigma \gsim 10^{-5}$ pb even for
$\tan\beta=70$ and $m_{H^+} \lsim 600$ GeV, we can conclude that,
in the context of the 2HDM-III, it is likely that a charged Higgs
boson could be observed at LHC energies by exploiting direct production.

In Fig.~\ref{fig:2}, we present results for the total cross
section rates of process $h_1 h_2 (c\bar{b}) \to H^+ X$ as a
function of $m_{H^+}$ in the 2HDM-II at LHC energies ($\sqrt s=14$ TeV),
by taking $(V_{\rm CKM})_{23}=4.16 \times 10^{-2}$ and $(V_{\rm CKM})_{33}
\approx 1$, for the cases: (a) $\tan\beta=0.1$, (b) $\tan\beta=1$,
(c) $\tan\beta=15$, and (d) $\tan\beta=70$. As we have already
said, the expected integrated luminosity at LHC is of the order
$10^5$ pb$^{-1}$, hence we also conclude from this figure that
 in the framework of the 2HDM-II we obtain production rates
for the charged Higgs boson { via} $c\bar b$-fusion that may be
detectable at LHC energies.

\subsection{Indirect production of charged Higgs bosons at the LHC}

We have found that, in some of the 2HDM-III scenarios
envisaged here, light charged Higgs bosons could exist that have
not been excluded by current experimental bounds, chiefly from LEP2
and Tevatron. Their discovery potential should therefore be studied
in view of the upcoming LHC and we shall then turn our attention now to
presenting the corresponding hadro-production cross sections via an
indirect channel, i.e., other than as secondary products in (anti)top
quark decays and via $c\bar b$-fusion, considered previously.

As dealt with so far, if the charged Higgs boson mass $m_{H^\pm}$
satisfies $m_{H^\pm} < m_{t} - m_{b}$, where $ m_{t}$ is the top
quark mass and $ m_{b}$ the bottom quark mass, $H^\pm$ particles
could be produced in the decay of on-shell (i.e., $\Gamma_t\to0$)
top (anti-)quarks $ t \rightarrow b H^+$ and the c.c. process, the latter being in turn produced in
pairs via $q\bar q$ annihilation and $gg$ fusion. We denote such a
$H^\pm$ production channel as $q\bar q$, $ gg \rightarrow t\bar t
\rightarrow t\bar bH^-$ + c.c.  (i.e., if due to (anti-)top
decays) whilst we use the notation $ q\bar q$, $ gg \rightarrow
t\bar bH^-$ + c.c.  to signify when further production diagrams
are included\footnote{Altogether, they represent the full gauge
invariant set of Feynman graphs pertaining to the $2\to3$ body
process with a $t\bar bH^-_i$ + c.c. final state: two for the case
of $q\bar q$ annihilation and eight for gluon-gluon fusion, see,
e.g., Eq.~(1.1) of \cite{Guchait:2001pi}.}. In fact, owing to the
large top decay width ($ \Gamma_{t} \geq 1.5$~GeV) and due to
the additional diagrams which do not proceed via direct $ t\bar t$
production but yield the same final state $t\bar bH^-$ + c.c.
\cite{Borzumati:1999th,Miller:1999bm,Moretti:1999bw}, charged
Higgs bosons could also be produced at and beyond the kinematic
top decay threshold. The importance of these effects in the
so-called `threshold' or `transition'  region ($m_{H^\pm}\approx
m_t$) was emphasized in various Les Houches
proceedings~\cite{Cavalli:2002vs,Assamagan:2004mu} as well as in
Refs.~\cite{Alwall:2003tc,Guchait:2001pi,Moretti:2002ht,Assamagan:2004gv},
so that the calculations of
Refs.~\cite{Borzumati:1999th,Miller:1999bm} (based on the
appropriate $q\bar q,gg\to tb H^\pm$ description) are now
implemented in
HERWIG\,\cite{herwig,Corcella:2000bw,Corcella:2002jc,Moretti:2002eu}\,and
PYTHIA\,\cite{pythia,Alwall:2004xw}. A comparison between the two
generators was carried out in Ref.~\cite{Alwall:2003tc}. For any
realistic simulation of $H^\pm$ production with $m_{H^\pm}\gsim
m_t$, as can well be the case here, the use of either of these two
implementations is of paramount importance.

\begin{figure}
\centering
\includegraphics[width=5in]{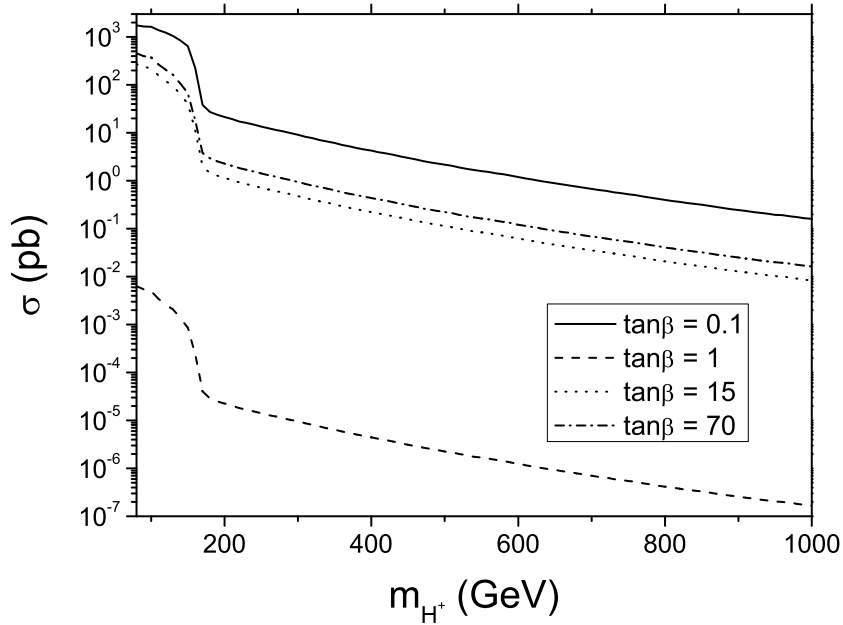}
\caption{ The figure shows the cross sections of $H^+$ production at
the LHC through the channel $q\bar q,gg\to t\bar b H^-$ + c.c. in
Scenario A ($\tilde{\chi}_{ij}^u = 1$ and $\tilde{\chi}_{ij}^d = 1$) and for $\tan \beta= 0.1,\, 1,\, 15\,, 70 $.} \label{fig:Xsec-a1}
\end{figure}

\begin{figure}
\centering
\includegraphics[width=5in]{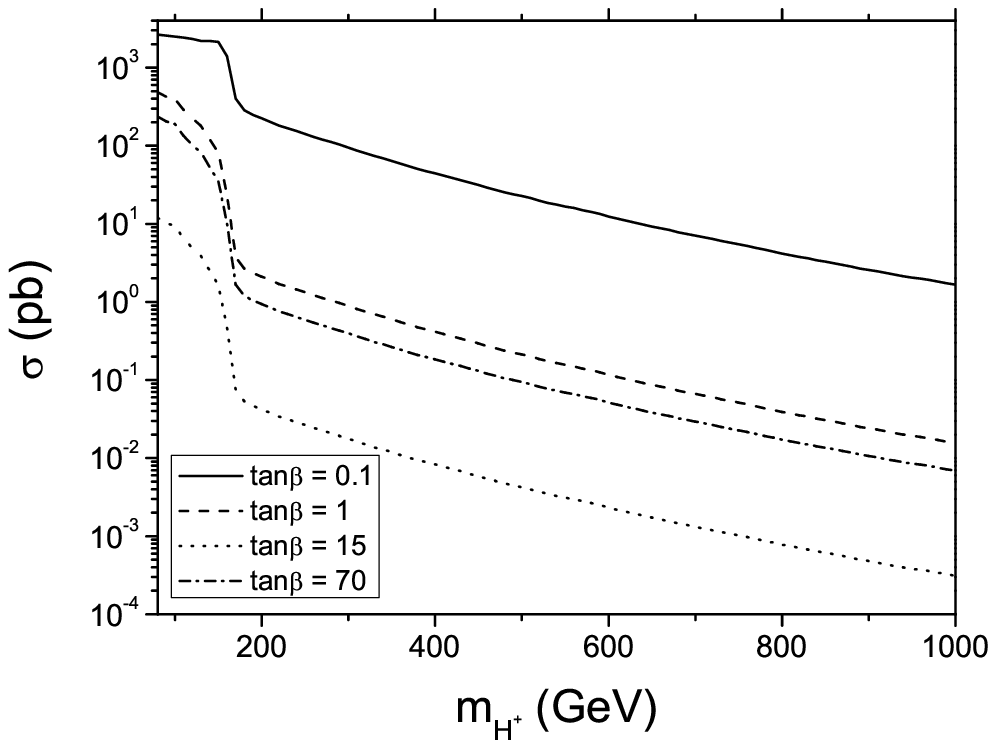}
\caption{ The same as in Fig.~\ref{fig:Xsec-a1} but taking
$\tilde{\chi}_{ij}^u = 1$ and $\tilde{\chi}_{ij}^d = 0.1$   (Scenario B).} \label{fig:Xsec-a2}
\end{figure}

\begin{figure}
\centering
\includegraphics[width=5in]{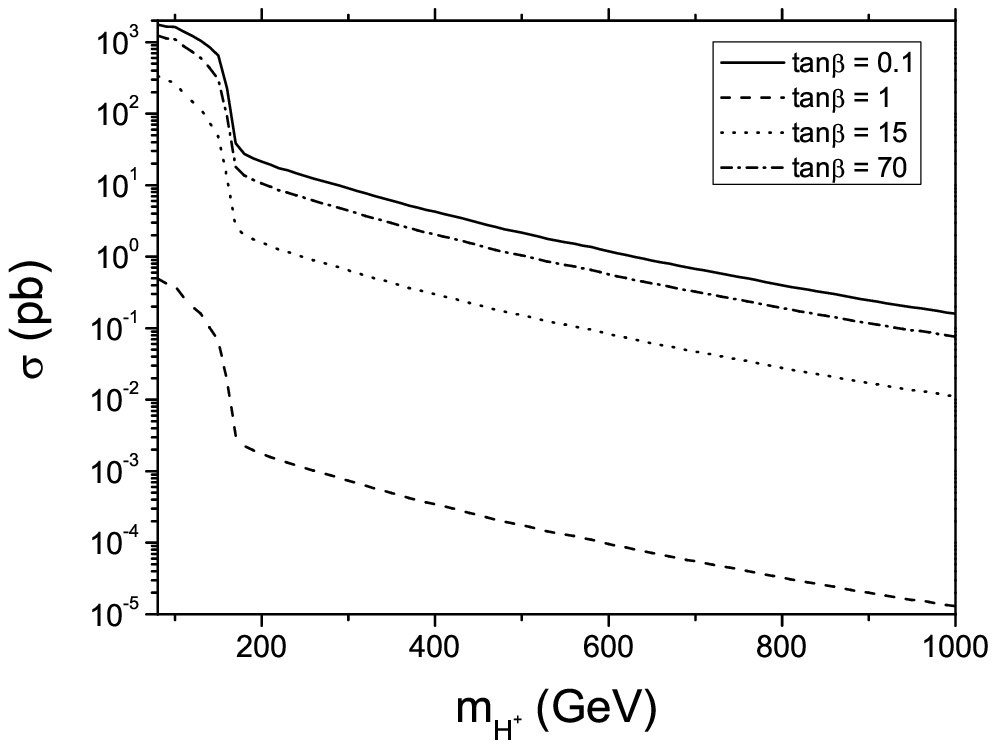}
\caption{ The same as in Fig.~\ref{fig:Xsec-a1} but taking
$\tilde{\chi}_{ij}^u = 0.1$ and $\tilde{\chi}_{ij}^d = 1$ (Scenario C).} \label{fig:Xsec-a3}
\end{figure}

\begin{figure}
\centering
\includegraphics[width=5in]{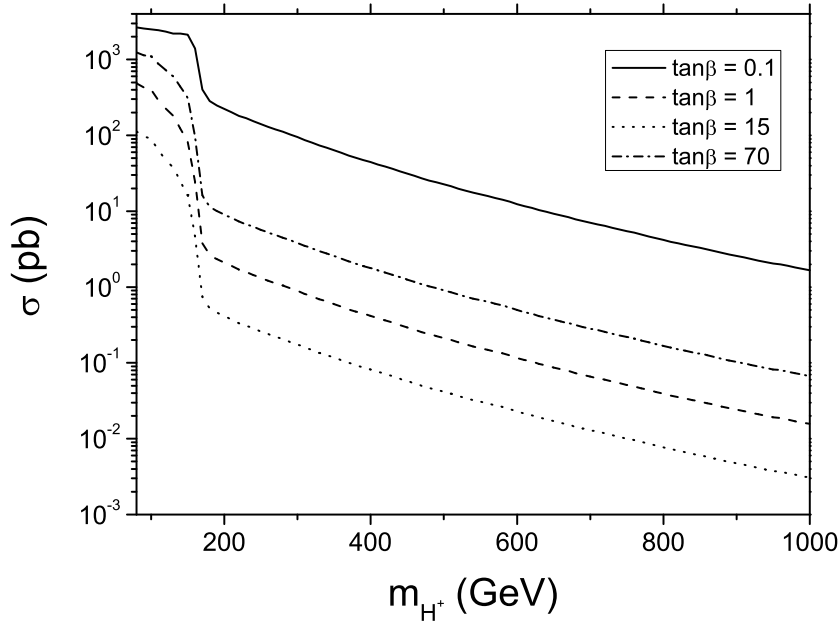}
\caption{ The same as in Fig.~\ref{fig:Xsec-a1} but taking
$\tilde{\chi}_{ij}^u = 0.1$ and $\tilde{\chi}_{ij}^d = 0.1$  (Scenario D).} \label{fig:Xsec-b1}
\end{figure}

Here, we use HERWIG version 6.510 in default configuration, by
onsetting the subprocess {\tt IPROC~=~3839}, wherein we have
overwritten the default MSSM/2HDM couplings and masses with those
pertaining to the 2HDM-III: see Eqs.~(\ref{coups1})--(\ref{sp}).
The production cross sections are found in
Figs.~\ref{fig:Xsec-a1}--\ref{fig:Xsec-b1}  for our usual
scenarios: A ($\tilde{\chi}_{ij}^u = 1$ and $\tilde{\chi}_{ij}^d =
1$), B ($\tilde{\chi}_{ij}^u = 1$ and $\tilde{\chi}_{ij}^d =
0.1$), C ($\tilde{\chi}_{ij}^u = 0.1$ and $\tilde{\chi}_{ij}^d =
1$) and D ($\tilde{\chi}_{ij}^u = 0.1$ and $\tilde{\chi}_{ij}^d =
0.1$) As usual, we adopt our four choices of
$\tan \beta$.

Altogether, by comparing the  $q\bar q,gg\to t\bar b H^-_i$ + c.c.
cross sections herein with, e.g., those of the MSSM in
\cite{Djouadi:2005gj} or the 2HDM in
\cite{Moretti:2002ht,Moretti:2001pp}, it is clear that the 2HDM-III
rates can be very large and thus the discovery potential in ATLAS
and CMS can be substantial, particularly for a very light
$H^\pm$, which may pertain to our 2HDM-III but not the MSSM or
2HDM-II. However, it is only by combining the production rates of
this section with the decay ones of the previous ones that actual
event numbers at the LHC can be predicted.

\section{Event rates of charged Higgs bosons at the LHC}

To illustrate the type of charged Higgs signatures that have the
potential to be detectable at the LHC in the 2HDM-III, we show in
Tabs.~\ref{tab:5} and \ref{tab:6}  the event rates of charged Higgs
boson through the channels $q\bar q,gg\to t\bar b H^-_i$ + c.c. and
$c\bar{b} \to H^+$ + c.c., alongside the corresponding production
cross sections ($\sigma$'s) and relevant BRs, for a combination of
masses, $\tan\beta$ and specific 2HDM-III parameters amongst those
used in the previous sections (assuming $m_{h^0}= 120$ GeV,
$m_{A^0}=300$ GeV and the mixing angle at $\alpha = \pi /2$
throughout). In particular, we focus on those cases where the
charged Higgs boson mass is above the threshold for $t \to b H^+$,
for two reasons. On the one hand, the scope of the LHC in accessing
$t \to b H^+$ decays has been established in a rather model
independent way. On the other hand, we have dealt at length with the
corresponding BRs in section III. (As default, we also assume an
integrated luminosity of $10^5$ pb$^{-1}$.)

To illustrate these results, let us comment on one case within each
scenario. From Table \ref{tab:5}, we can see that for Scenario A,
with $(\tilde{\chi}_{ij}^{u}=1, \tilde{\chi}_{ij}^{d}=1)$ and
$\tan\beta=15$, we have that the ${H^{\pm}}$ is heavier than $m_t -m_b$,
as we take a mass $m_{H^+}=400$, thus precluding top decay
contributions, so that in this case $\sigma(pp \to t \bar{b} H^+)
\approx 2.2 \times 10^{-1}$ pb, while the dominant decays are $H^+
\to t \bar{b}, \tau^+ \nu_\tau \, W^+ h^0,  \, W^+ A^0$ which give
a number of events of 7040, 46, 13860, 374, respectively. In this
case the most promising signal is $H^+ \to W^+ h^0$. However, when
$\tan \beta=70$ we have that all event rates increase substantially. Here,
the signal $H^+ \to W^+ h^0$ is still the most important with an
event rate of 15480.

Then, for Scenario B $(\tilde{\chi}_{ij}^{u}=0.1,
\tilde{\chi}_{ij}^{d}=1)$, we have that $H^\pm$ is again above the
threshold for $t \to H^+ b$. So, for the declared values of the
relevant parameters, we take a charged Higgs boson mass
$m_{H^+}=600$ for $\tan \beta=1$ and $\tan \beta=70$,
respectively. In such a case  the decay $H^+ \to W^+ h^0$ can
reach significant numbers for the LHC. We obtain a number of
events of 3960 and 2703, respectively. The other decay that has a
large BR is $H^+ \to W^+ A^0$, and in these cases the number of
events ranges over 1200--3500.

Next, we discuss the Scenario C $(\tilde{\chi}_{ij}^{u}=1,
\tilde{\chi}_{ij}^{d}=0.1)$ for $\tan \beta =15$. Here, we obtain
that the signals $H^+ \to t \bar{b}$ and $W^+ h^0$ are the most
relevant ones, with a number of events about 34560 and 26240, respectively.

Finally, for scenario D
$(\tilde{\chi}_{ij}^{u}=0.11, \tilde{\chi}_{ij}^{d}=0.1)$ the
dominant decays are $H^+ \to t \bar{b}, \tau^+ \nu_\tau$ and $ W^+
h^0$, which give a spectacular number of events: 269800, 68400
and 34200, respectively. Here, we have set $\tan\beta=70$.

All these rates correspond to the case of indirect production. The
contribution due to direct production is in fact subleading,
especially at large $m_{H^\pm}$ values. Nonetheless, in some
benchmark cases, they could represent a sizable addition to the
signal event rates. This is especially the case for Scenario A with
$\tan\beta=15$ or 70 and Scenario C with $\tan\beta=15$. In general
though, also considering the absence of an accompanying trigger
alongside the $H^\pm$,{\it i.e.} for instance a top quark produced
in $g b \to H^- t$ could help to identify the signal. Thus, we
expect that the impact of $c\bar b$-fusion at the LHC will be more
marginal that that of $gg$-fusion for large Higgs masses, in fact,
at times even smaller that the contribution from $q\bar
q$-annihilation.

\squeezetable
\begin{table*}[htdp]
\caption{\label{tab:5} Summary of LHC event rates for some parameter combinations within Scenarios A, B, C, D with
  for an integrated luminosity of $10^{5}$
pb$^{-1}$, for several different signatures, through the channel $q
\bar q,gg \to \bar{t} b H^+$ + c.c.}

\begin{tabular}{|c|c|c|c|c|c|}
\hline $(\tilde{\chi}_{ij}^{u}, \tilde{\chi}_{ij}^{d})$  & $\tan\beta$ & $m_{H^+}$ in GeV &
$\sigma(pp \to H^+ \bar{t} b)$ in pb & Relevant BRs & Nr. Events\\

\hline \multicolumn{1}{|c|}{ (1,1)  }& 15 &400 &
$2.23\times 10^{-1}$ &\begin{tabular}{l} BR$\left( H^{+} \to
t\bar{b}\right)\approx 3.2 \times 10^{-1} $\\
BR$\left( H^{+} \to
\tau^{+}\nu^{0}_\tau\right) \approx 2.1 \times 10^{-3} $\\
BR$\left( H^{+} \to
W^{+}h^{0}\right) \approx 6.3 \times 10^{-1} $\\
BR$\left( H^{+}_{2} \to
W^{+}A^{0}\right) \approx 1.7 \times 10^{-2} $%
\end{tabular}
& \multicolumn{1}{|c|}{$
\begin{tabular}{r}
7040\\
46\\
13860\\
374%
\end{tabular}
$} \\ \hline

\hline \multicolumn{1}{|c|}{ (1,1) }& 70 & 400 &
$4.3\times 10^{-1}$ &\begin{tabular}{l} BR$\left( H^{+} \to
t\bar{b}\right)\approx 3.5 \times 10^{-1} $\\
BR$\left( H^{+} \to
c \bar{b}\right) \approx 1.4 \times 10^{-2} $\\
BR$\left( H^{+} \to
\tau^{+}\nu_{\tau}\right) \approx 2.5 \times 10^{-1} $\\
BR$\left( H^{+} \to
W^{+}h^{0} \right) \approx 3.6 \times 10^{-1} $%
\end{tabular}
& \multicolumn{1}{|c|}{$
\begin{tabular}{r}
15050\\
602\\
10750\\
15480%
\end{tabular}
$} \\ \hline

\hline \multicolumn{1}{|c|}{ (0.1,1) }& 1 & 600 &
$1.1\times 10^{-1}$ &\begin{tabular}{l} BR$\left( H^{+} \to
t\bar{b} \right) \approx 3 \times 10^{-1} $\\
BR$\left( H^{+} \to
t\bar{s} \right)\approx 9.1 \times 10^{-4} $\\
BR$\left( H^{+} \to
W^{+}h^{0}\right) \approx 3.6\times 10^{-1} $\\
BR$\left( H^{+} \to
W^{+}A^{0}\right) \approx 3.2 \times 10^{-1} $%
\end{tabular}
& \multicolumn{1}{|c|}{$
\begin{tabular}{r}
3300\\
10\\
3960\\
3520%
\end{tabular}
$} \\ \hline

\hline \multicolumn{1}{|c|}{ (0.1,1)}& 70 & 600 &
$5.1\times 10^{-2}$ &\begin{tabular}{l} BR$\left( H^{+} \to
\tau^+\nu_\tau\right)\approx 1.2 \times 10^{-1} $\\
BR$\left( H^{+} \to
t\bar{b}  \right)\approx 9.4 \times 10^{-2} $\\
BR$\left( H^{+} \to
W^{+}h^{0}\right) \approx 5.3 \times 10^{-1} $\\
BR$\left( H^{+} \to W^{+}A^{0}\right) \approx 2.3 \times
10^{-1} $%
\end{tabular}
& \multicolumn{1}{|c|}{$
\begin{tabular}{r}
612\\
470\\
2703\\
1173%
\end{tabular}
$} \\ \hline

\hline \multicolumn{1}{|c|}{ (1,0.1) }& 15 &300 &
$6.4\times 10^{-1}$ &\begin{tabular}{l} BR$\left( H^{+} \to
t\bar{b} \right)\approx 5.4 \times 10^{-1} $\\
BR$\left( H^{+} \to
c\bar{b} \right) \approx 5 \times 10^{-4} $\\
BR$\left( H^{+} \to
\tau^+\nu_\tau \right) \approx 3.9 \times 10^{-2} $\\
BR$\left( H^{+} \to
W^{+}h^{0}\right) \approx 4.1 \times 10^{-1} $%
\end{tabular}
& \multicolumn{1}{|c|}{$
\begin{tabular}{r}
34560\\
32\\
2535\\
26240%
\end{tabular}
$} \\ \hline

\hline \multicolumn{1}{|c|}{ (0.1,0.1) }& 70 & 300 &
$3.8 $ &\begin{tabular}{l} BR$\left( H^{+} \to
\tau^+\nu_\tau\right)\approx 1.8 \times 10^{-1} $\\
BR$\left( H^{+} \to
t\bar{b}\right)\approx 7.1 \times 10^{-1} $\\
BR$\left( H^{+} \to
c\bar{b} \right) \approx 2.4 \times 10^{-3} $\\
BR$\left( H^{+} \to
W^{+}h^{0}\right) \approx 9 \times 10^{-2} $%
\end{tabular}
& \multicolumn{1}{|c|}{$
\begin{tabular}{r}
68400\\
269800\\
912\\
34200%
\end{tabular}
$} \\ \hline

\end{tabular}
\label{default5}
\end{table*}
\squeezetable
\begin{table*}[htdp]
\caption{\label{tab:6} Summary of LHC event rates for some parameter combinations within Scenarios A, B, C, D with
  for an integrated luminosity of $10^{5}$
pb$^{-1}$, for several different signatures, through the channel $c
\bar b \to H^+$ + c.c.}

\begin{tabular}{|c|c|c|c|c|c|}
\hline $(\tilde{\chi}_{ij}^{u}, \tilde{\chi}_{ij}^{d})$  & $\tan\beta$ & $m_{H^+}$ in GeV &
$\sigma(pp \to H^+ + X)$ in pb & Relevant BRs & Nr. Events\\

\hline \multicolumn{1}{|c|}{ (1,1) }& 15 &400 & $ 1.14 \times
10^{-1}$ &\begin{tabular}{l} BR$\left( H^{+} \to
t\bar{b}\right)\approx 3.2 \times 10^{-1} $\\
BR$\left( H^{+} \to
\tau^{+}\nu^{0}_\tau\right) \approx 2.1 \times 10^{-3} $\\
BR$\left( H^{+} \to
W^{+}h^{0}\right) \approx 6.3 \times 10^{-1} $\\
BR$\left( H^{+}_{2} \to
W^{+}A^{0}\right) \approx 1.7 \times 10^{-2} $%
\end{tabular}
& \multicolumn{1}{|c|}{$
\begin{tabular}{r}
3648\\
24\\
7182\\
194%
\end{tabular}
$} \\ \hline

\hline \multicolumn{1}{|c|}{ (1,1) }& 70 & 400 & $1.25\times
10^{-1}$ &\begin{tabular}{l} BR$\left( H^{+} \to
t\bar{b}\right)\approx 3.5 \times 10^{-1} $\\
BR$\left( H^{+} \to
c \bar{b}\right) \approx 1.4 \times 10^{-2} $\\
BR$\left( H^{+} \to
\tau^{+}\nu_{\tau}\right) \approx 2.5 \times 10^{-1} $\\
BR$\left( H^{+} \to
W^{+}h^{0} \right) \approx 3.6 \times 10^{-1} $%
\end{tabular}
& \multicolumn{1}{|c|}{$
\begin{tabular}{r}
4375\\
175\\
3125\\
4500%
\end{tabular}
$} \\ \hline

\hline \multicolumn{1}{|c|}{ (0.1,1) }& 1 & 600 & $3.41\times
10^{-4}$ &\begin{tabular}{l} BR$\left( H^{+} \to
t\bar{b} \right) \approx 3 \times 10^{-1} $\\
BR$\left( H^{+} \to
t\bar{s} \right)\approx 9.1 \times 10^{-4} $\\
BR$\left( H^{+} \to
W^{+}h^{0}\right) \approx 3.6\times 10^{-1} $\\
BR$\left( H^{+} \to
W^{+}A^{0}\right) \approx 3.2 \times 10^{-1} $%
\end{tabular}
& \multicolumn{1}{|c|}{$
\begin{tabular}{r}
10\\
0\\
12\\
11%
\end{tabular}
$} \\ \hline

\hline \multicolumn{1}{|c|}{ (0.1,1)}& 70 & 600 & $1.98\times
10^{-3}$ &\begin{tabular}{l} BR$\left( H^{+} \to
\tau^+\nu_\tau\right)\approx 1.2 \times 10^{-1} $\\
BR$\left( H^{+} \to
t\bar{b}  \right)\approx 9.4 \times 10^{-2} $\\
BR$\left( H^{+} \to
W^{+}h^{0}\right) \approx 5.3 \times 10^{-1} $\\
BR$\left( H^{+} \to W^{+}A^{0}\right) \approx 2.3 \times
10^{-1} $%
\end{tabular}
& \multicolumn{1}{|c|}{$
\begin{tabular}{r}
24\\
19\\
105\\
45%
\end{tabular}
$} \\ \hline

\hline \multicolumn{1}{|c|}{ (1,0.1) }& 15 &300 & $3.99\times
10^{-1}$ &\begin{tabular}{l} BR$\left( H^{+} \to
t\bar{b} \right)\approx 5.4 \times 10^{-1} $\\
BR$\left( H^{+} \to
c\bar{b} \right) \approx 5 \times 10^{-4} $\\
BR$\left( H^{+} \to
\tau^+\nu_\tau \right) \approx 3.9 \times 10^{-2} $\\
BR$\left( H^{+} \to
W^{+}h^{0}\right) \approx 4.1 \times 10^{-1} $%
\end{tabular}
& \multicolumn{1}{|c|}{$
\begin{tabular}{r}
21546\\
20\\
1556\\
16359%
\end{tabular}
$} \\ \hline

\hline \multicolumn{1}{|c|}{ (0.1,0.1) }& 70 & 300 & $3.88 \times
10^{-1} $ &\begin{tabular}{l} BR$\left( H^{+} \to
\tau^+\nu_\tau\right)\approx 1.8 \times 10^{-1} $\\
BR$\left( H^{+} \to
t\bar{b}\right)\approx 7.1 \times 10^{-1} $\\
BR$\left( H^{+} \to
c\bar{b} \right) \approx 2.4 \times 10^{-3} $\\
BR$\left( H^{+} \to
W^{+}h^{0}\right) \approx 9 \times 10^{-2} $%
\end{tabular}
& \multicolumn{1}{|c|}{$
\begin{tabular}{r}
6984\\
27548\\
93\\
3492%
\end{tabular}
$} \\ \hline

\end{tabular}
\label{default6}
\end{table*}

\section{Conclusions}

We have discussed the implications of assuming a four-zero Yukawa
texture for the properties of the charged Higgs boson, within the
context of a 2HDM-III. In particular, we have presented a detailed
discussion of the charged Higgs boson couplings to heavy fermions
and the resulting pattern for its decays. The latter clearly
reflect the different coupling structure of the 2HDM-III, e.g.,
with respect to the 2HDM-II, so that one has at disposal more
possibilities to search for $H^{\pm}$ states at current and future
colliders, ideally enabling one to distinguish between different
Higgs models of EWSB. We have then concentrated our analysis to
the case of the LHC and showed that the production rates of
charged Higgs bosons at the LHC is sensitive to the modifications
of the Higgs boson couplings. We have done so by evaluating
2HDM-III effects on the top decay $t \to b H^{+}$ as well as in
the $s$-channel production of $H^{\pm}$ through $c\bar{b}$-fusion
and the multibody final state induced by $gg$-fusion and $q\bar
q$-annihilation. Finally, we have determined the number of events
for the most promising LHC signatures of a $H^\pm$ belonging to a
2HDM-III, for both $c \bar{b}\to H^+$ + c.c. and $q\bar q\to \bar
t bH^+$ + c.c. scatterings (the latter affording larger rates than
the former). Armed with these results, we are now in a position to
carry out a detailed study of signal and background rates, in
order to determine the precise detectability level of each
signature. However, this is beyond the scope of present work and
will be the subject of a future publication.

\bigskip

\section*{Acknowledgements}
 This work was supported in part by CONACyT and SNI
(M\'exico). J.H-S. thanks in particular CONACyT (M\'exico) for the
grant J50027-F and SEP (M\'exico) by the grant PROMEP/103.5/08/1640.
R.N-P. acknowledges the Institute of Physics BUAP for a warm
hospitality and also financial support by CONACyT through the
program \emph{Apoyo Complementario para la Consolidaci\'on
Institucional de Grupos de Investigaci\'on (Retenci\'on)}.

\end{document}